\documentclass[]{aastex631}

\usepackage{color, xcolor}
\usepackage{soul}
\soulregister\cite7
\soulregister\citep7

\begin{document}

\title{Taxonomic analysis of asteroids with artificial neural networks}

\author{Nanping Luo}
\affiliation{Yunnan Observatories, CAS, \\
Kunming, 650216, China \\
}
\affiliation{University of Chinese Academy of Sciences, Beijing 100049, China\\
}

\author{Xiaobin Wang}\thanks{E-mails: wangxb@ynao.ac.cn;luonanping@ynao.ac.cn}
\affiliation{Yunnan Observatories, CAS, \\
Kunming, 650216, China \\
}
\affiliation{University of Chinese Academy of Sciences, Beijing 100049, China\\
}
\affiliation{Key Laboratory for the Structure and Evolution of Celestial Objects, Chinese Academy of Sciences, Kunming 650216, China\\
}

\author{Shenghong Gu}
\affiliation{Yunnan Observatories, CAS, \\
Kunming, 650216, China \\
}
\affiliation{University of Chinese Academy of Sciences, Beijing 100049, China\\
}
\affiliation{Key Laboratory for the Structure and Evolution of Celestial Objects, Chinese Academy of Sciences, Kunming 650216, China\\
}

\author{Antti Penttilä}
\affiliation{Department of Physics, P.O. box 64, FI-00014 University of Helsinki, Finland}

\author{Karri Muinonen}
\affiliation{Department of Physics, P.O. box 64, FI-00014 University of Helsinki, Finland}

\author{Yisi Liu}
\affiliation{ Deep Space Exploration Laboratory, Beijing 100043, China
} 

\begin{abstract}

We study the surface composition of asteroids with visible and/or infrared spectroscopy. For example, asteroid taxonomy is based on the spectral features or multiple color indices in visible and near-infrared wavelengths. The composition of asteroids gives key information to understand their origin and evolution.  However, we lack compositional information for faint asteroids due to limits of ground-based observational instruments. In the near future, the Chinese Space Survey telescope (CSST) will provide multiple colors and spectroscopic data for asteroids of apparent magnitude brighter than 25 mag and 23 mag, respectively.  For the aim of analysis of the CSST spectroscopic data, we applied an algorithm using artificial neural networks (ANNs) to establish a preliminary classification model for asteroid taxonomy according to the design of the survey module of CSST.  Using the SMASS II spectra and the Bus-Binzel taxonomy system, our ANN classification tool composed of 5 individual ANNs is constructed, and the accuracy of this classification system is higher than 92 \%.  As the first application of our ANN tool, 64 spectra of 42 asteroids obtained in 2006 and 2007 by us with the 2.16-m telescope in the Xinglong station (Observatory Code 327) of National Astronomical Observatory of China are analyzed. The predicted labels of these spectra using our ANN tool are found to be reasonable when compared to their known taxonomic labels. Considering the accuracy and stability, our ANN tool can be applied to analyse  the CSST asteroid spectra in the future.
\end{abstract}


\section{Introduction} \label{sec:intro}

Small Solar System objects S3Os are thought to be remnants of planetesimals from the early stage of the planetary formation of the Solar System. Compared to the planets, the S3Os  could retain more information of protoplanetary conditions because of suffering less secondary chemical and geological evolution, although they have undergone collisions, space weathering, and dynamical and thermal evolution, which shaped their present physical and orbital properties \citep{2014Natur.505..629D}. At present, most discovered S3Os are asteroids which are thought to originate from the inner planetesimals, as the building blocks of the terrestrial planets. The composition of asteroids vs. their orbits can provide some clues to the origin and the evolution of asteroids, as well as the constraints on planetary formation models in the inner Solar System \citep{2002aste.book.....B}.

The surface composition of asteroids can be inferred from photometric colors and/or spectroscopic data in the visual (Vis) and near-infrared (NIR) wavelength regimes. The overall shape and absorption features of reflectance spectra of asteroids reflect the compositions and minerals they contain. Additionally, there are four effects that affect the reflectance spectra of asteroids: (1) phase reddening, which raises the spectral slope by 8--12 \% in the phase angle range of 0°--100° \citep{2002NEAR};  (2) space weathering,  which makes a planetary surface darken and redden; (3) the particle size distribution of the regolith, which changes spectral slope and band depth;  and (4) the temperature of the surface, which affects the shapes of spectral bands, e.g., associated with olivine and pyroxene minerals. The spectral taxonomy is a frequently used method to understand the surface composition of S3Os, and is based on the spectral slope and absorption features in the spectra. With the contributions of many astronomers, several classification systems of asteroids have been established. Using the data of the Eight-Color Asteroid Survey (ECAS, the wavelength range is 0.337--1.041 \textmu{}m), the Tholen taxonomy classified asteroids into 14 classes \citep{1985Icar...61..355Z}. The SMASS II taxonomy, based on the data of the Small Main-Belt Asteroid Spectroscopic Survey in 2002 (SMASS II, the wavelength range of 0.435--0.925 \textmu{}m), classified asteroids into 26 types \citep{2002Icar..158..146B}. The Bus-DeMeo taxonomy \citep{2009Icar..202..160D} classified asteroids into 24 classes using both visible and near-infrared data (the wavelength range of 0.45--2.45 \textmu{}m).

Recently, taxonomic studies of asteroids have started to apply machine learning techniques to, e.g., multiple photometric colors \citep{2022arXiv220405075C,2018A&A...617A..12P}, spectra \citep{2021A&A...649A..46P}, albedo \citep{2017Icar..284...30B,2022arXiv220311229M}, and spectral slope  \citep{1985Icar...61..355Z,2009Icar..202..160D}. Various machine learning algorithms were applied to asteroid taxonomy, for example, principal component analysis \citep[PCA]{2002Icar..158..146B, 2009Icar..202..160D}, random forest \citep[RF]{2017ChA&A..41..549H}, cluster analysis \citep[CA]{2022arXiv220311229M}, multinomial logistic regression \citep[MLR]{2021FrASS...8..216K}, naive Bayes \citep{2021FrASS...8..216K}, support vector machines \citep[SVM]{2021FrASS...8..216K}, gradient boosting \citep[GB]{2021FrASS...8..216K}, multilayer perceptron (MLP), or feed-forward neural network \citep{2021FrASS...8..216K,2021A&A...649A..46P,1994JGR....9910847H}. The PCA method refers to a transformation of a dataset of multiple dimensions, which could reduce the dimensionality of the data and reach a goal that the information of data set concentrated on a few variables \citep{2019arXiv190407248B,2019sdmm.book.....I,2010IJMPD..19.1049B}.  The RF method is composed of decision trees, each decision tree giving a classification model. Repeatedly constructing decision trees, the RF outputs the final results of classification by voting process among the constructed decision trees \citep{2019arXiv190407248B,2019sdmm.book.....I,2010IJMPD..19.1049B,2017ChA&A..41..549H}. MLP is the most used model among the artificial neural networks (ANNs), and usually applies a back-propagation algorithm (BP).  Comparatively, the feed-forward neural network can have a higher accuracy when used to classify the spectroscopic data of asteroids \citep{2021FrASS...8..216K}. Such a neural network has a good nonlinear function approximation performance and prediction ability. It is robust to irrelevant or redundant attributes, which makes it able to adapt to the complex data environment with various features \citep{2019arXiv190407248B,2019sdmm.book.....I,2010IJMPD..19.1049B}.
In the analysis of asteroid taxonomy, the neural network techniques could make full use of the spectral trend and absorption features, especially of those that are difficult to distinguish by human eye. The ANN is a supervised machine learning algorithm and has been applied successfully in the analysis of asteroid taxonomy. For example, \citet{2021A&A...649A..46P} applied the algorithm to build a neural network of 11 classes, reduced from the Bus-DeMeo taxonomic system \citep{2009Icar..202..160D}, with a combination dataset using Vis-NIR wavelength range between 0.45 and 2.45 µm. \citet{2021A&A...649A..46P} showed the established ANN to work well in classifying asteroids, even in situations where the wavelength range of the tested spectra differs from that of the original taxonomic system, and concluded that the ANN could be applied to the spectral data of ESA Gaia space telescope.

In practice, most of the discovered asteroids have no spectral data due to their faint brightness. In the near future, the situation will be improved with large ground-based telescopes, i.e., the Vera C. Rubin Observatory, and space survey telescopes such as the CSST (Chinese Space Survey Telescope) and the ESA Euclid.

The ESA Euclid mission will employ a 1.2-m telescope to survey a sky area of 15,000 square degree in its six-years running period. Its Vis-NIR imaging and spectroscopic instruments could detect celestial objects of apparent magnitude brighter than $V_{AB}=24.5$~mag. Accordingly, roughly 150,000 S3Os (mainly main-belt asteroids) could be detected by the imaging mode of Euclid, and 100,000 could be spectroscopically observed in the wavelength range of 0.5--2 \textmu{}m.

\begin{figure*}[ht!]
\centering
\includegraphics[width=0.5\columnwidth]{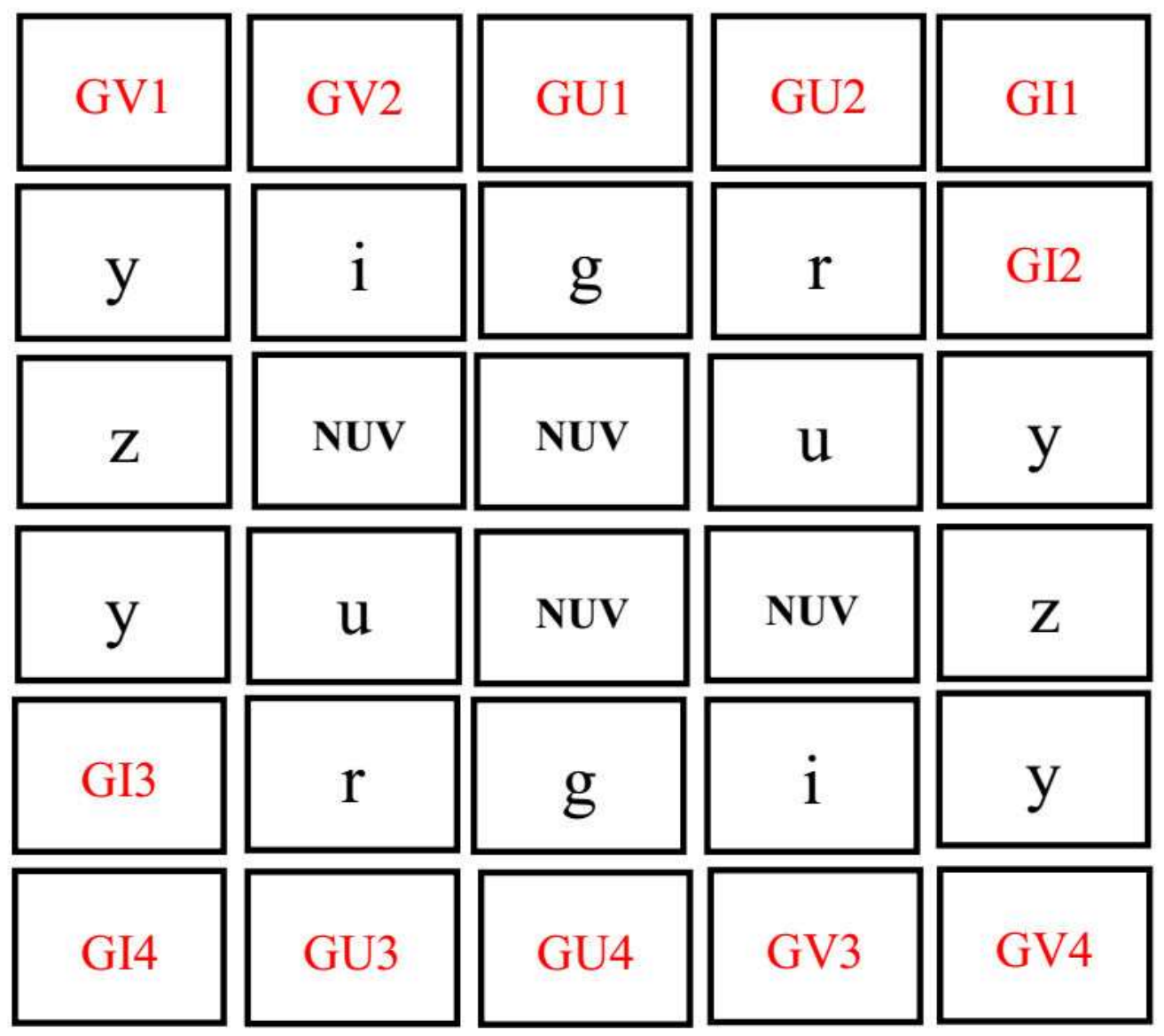}
\caption{The CCDs, filters, and gratings arrangement in the survey module of the CSST.
\label{fig:ccd}}
\end{figure*}

The CSST, a 2.0-m telescope with a field of view of $1.2\!\times\!1.2$ deg, is expected to be launched in 2024 and located next to the Chinese space station \citep{zhanhu}. At present, the CSST has 5 observational modules: Survey module, Terahertz module (THz), Multi-Channel Imager (MCI), Integrated Field of view Spectrometer (IFS), and Cool Planet Image star Coronagraph module (CPI-C). During its ten-year operation, the survey module will occupy 70 \% of the total observation time. Consequently, a 17,500 deg$^2$ deep-sky area and a 400 deg$^2$ extreme-deep sky area could be obtained. The exposure times for the deep-sky area and the extreme-deep sky area will be 150 s and 240 s, respectively. The survey module of the CSST is composed of 30 CCD detectors with imaging  and spectroscopy functions (see Fig.~\ref{fig:ccd}). The imaging part consists of 18 CCDs and 7 color filters: NUV (0.22--0.32 \textmu{}m), u (0.32--0.4 \textmu{}m), g (0.4--0.55 \textmu{}m), r (0.55--0.69 \textmu{}m), i (0.69--0.82 \textmu{}m), z (0.82--1.0 \textmu{}m), and possibly a near-infrared filter (0.9--1.7 \textmu{}m). The slitless spectroscopic part involves 12 CCDs and three gratings \citep{zhanhu}: GU (0.255--0.4 \textmu{}m), GV (0.4--0.6 \textmu{}m), and GI (0.6--1.0 \textmu{}m). From the designed three gratings, the spectral observations will cover a wavelength of 0.255 to 1.0 \textmu{}m. Roughly, the limiting magnitudes of imaging observations for the deep-sky and extreme-deep sky modes are 25.5 and 26.5 mag, respectively, and they are 23 and 24 mag for spectroscopic observations in the deep-sky and extreme-deep sky survey. According to the planned ten-year observations with the CSST's survey module and the averaged  magnitude limit, around 200,000 known S3Os  will be observed and new S3Os will be discovered in the survey observations.

In the near future, a largely extended spectral database of asteroids will be obtained by large ground-based telescopes and space-based telescopes. The taxonomic analysis of asteroids is necessary for these new spectral data in order to understand the origin and evolution of asteroids. To meet the goal, we developed a spectral analysis platform for the spectral data of asteroids obtained by the CSST survey module with the aid of an artificial neural network (ANN) technique.

To build the ANN for asteroid taxonomy to be applied in the  future to the CSST spectroscopic data, we used the SMASS II data \citep{2002Icar..158..146B} and  the Bus-Binzel taxonomy system \citep{2002Icar..158..106B}. The design of the spectroscopic units of the CSST's survey module is going to provide us with the spectral data of a wavelength range of 0.255 to 1.0 \textmu{}m with an rough resolution of 200 \citep{zhanhu}. Therefore, the SMASS II data of the wavelength range of 0.435 and 0.925 \textmu{}m, used as the training and test procedures of ANN, are re-sampled according to the resolution of the CSST's spectroscopic gratings.  As the application of the derived ANN,  64 spectra of 42 asteroids observed by us in 2006 and 2007 with the 2.16-m telescope in Xinglong site of the National Astronomical Observatory of China are analysed.

This paper is arranged as follows. Section \ref{sec:ann}  introduces the construction of the ANN applied in this work,  and Section \ref{sec:data} describes the preparation of the training and test data to be used with the ANN. The procedure of training and evaluation of the ANN is shown in Section \ref{sec:con}. As an application of our ANN, in Section \ref{sec:app}, the analysis results with the derived ANN are presented, and the final Section \ref{sec:sum} gives a summary of the  work.

\section{Artificial neural network} \label{sec:ann}

Machine learning techniques are frequently used to solve problems in the field of, e.g., classification and regression. The machine learning algorithms are usually divided into four types: supervised learning, unsupervised learning, self-supervised learning, and reinforcement learning. The ANN built for the taxonomic analysis of asteroids belongs to supervised learning, it "learns" by looking for the mapping relationship between the features of the input data to their corresponding labels. As a result, the trained ANN classification model giving a high accuracy for test data can be applied to new spectra of asteroids.

\subsection{Structure of the artificial neural network} \label{subsec:Structure }

\begin{figure}[ht!]
\plotone{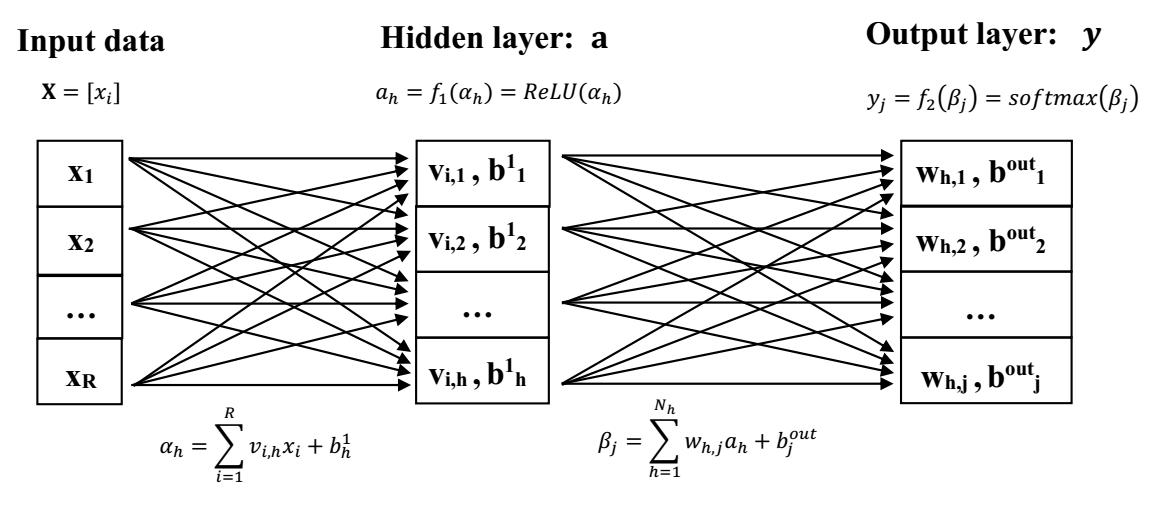}
\caption{The structure of the artificial neural network. The network contains an input layer, a hidden layer, and an output layer. $X$ represents the input, $v$ and $W$ represent weights, $b$ represents biases, $f$ represents the transfer function, and $y$ represents the output.
\label{fig:ann}}
\end{figure}

We constructed an ANN of an input layer, one hidden layer, and an output layer (see Fig.\ref{fig:ann}). Neurons in the input layer take the elements of an input data or features of spectra, represented by a vector $X=[x_i],\;i=1,\ldots,R$. The number of neurons in the input layer $R$ is the number of features in the input spectral data.  The $K$ neurons in the output layer present a discrete probability distribution $Y=[y_j],\;j=1,\ldots,K$ for $K$ classes ($j$ is the class index, and neuron $y_j$  in the output layer presents the probability of the input spectral data belonging to class $j$).

Here, we temporarily set the number of neurons in the input layer as $R = N_{in}= 151$ considering the wavelength range of available training data (i.e., SMASS II, 0.435 $\sim$ 0.925 \textmu{}m ), and a lower limitation of the resolution of 200 of the CSST slitless spectroscopic units. The concrete element $\lambda_i$ of input data of our ANNs is computed by Eq.(1), in which $\lambda_0$ is the shortest wavelength in the training data or test data, $(1+\frac{1}{R})^i$ ($i$, the index of element of input data) is the interval distance of two adjacent elements of $i$ and $i-1$. For a wavelength range of spectral data from 0.435 to 0.925 \textmu{}m, the maximum of $i$ is 151:
\begin{equation}
    \lambda_{i}=\lambda_{o} (1+\frac{1}{R})^{i}, \;\; (i=1,\ldots,151).
\end{equation}
The number of neurons in the output layer is set as $K=10$, namely, only ten most popular classes of asteroids are included in the presented ANNs. Detailed information about the chosen asteroid classes is described in Sec.~3. 

The number of neurons in the hidden layer is usually hard to determine accurately due to its flexible property. Often, it is determined by iterating over different choices. Generally, more neurons in the hidden layer allow more complicated problems to be solved by the ANN. One should take care when choosing the number of neurons in the hidden layer, because too few neurons in the hidden layer would result in under-fitting the input data, while too large number of neurons can lead to over-fitting. In other words, for the under-fitting case, the ANN will give a low accuracy of prediction to the input data, and for the over-fitting case, the ANN could give a high accuracy to training data, but a low accuracy to the test data. In our case, we found $N_h=45$ neurons in the hidden layer to work well.

As a feed-forward neural network, the input information is transferred to neurons in the hidden layer, and the output of the hidden layer is continued to be transferred to the neurons of the output layer, and the output of the output layer offers us the results.  The transfer, or the connections between the neurons in two consecutive layers, is done using weights, biases, and activation functions. 

In detail, the input layer receives the  information/data by storing the features/elements of input in the neurons.  Each neuron $h$ in the hidden layer receives information/signals from all neurons $[x_i],i=1,N_{in}$ in the input layer, multiplies that with the weights $v_{i,h}$ and adds bias $b^1_{h}$,  finally outputting $a_h$ after a nonlinear activation function (see Eq. (2)). In here we use the ReLU function which can improve the convergence rate of the estimation of weights and biases:
\begin{equation}
\begin{array}{rl}
     \alpha_h&=\sum\limits_{i=1}^{N_{in}}v_{i,h}x_i+b^1_h\\
    a_h=f_1(\alpha_h)=&{\rm ReLU}(\alpha_h)=\max(\alpha_h,0) \\
    \end{array}
\end{equation}

The output of the hidden layer $a_h, h=1,\ldots,N_{h}$ is taken as the input of the output layer. With a similar procedure but using a different activation function, the information continue to be passed to the neurons in the  output layer with the weights $w_{h,j}$ and the biases $b^{out}_j$. With a classification problem as in our work, the so-called softmax function is taken as the activation function, denoted as $f_2$, between the hidden layer and the output layer. After the transfer function $f_2$, the output of the output layer $Y=[y_j\epsilon[0,1]],\;j=1,\ldots,K$ (see Eq.(3)) is the predicted probability distribution for the classes: 
\begin{equation}
\begin{array}{rl}
     \beta_j&=\sum\limits_{h=1}^{N_h}w_{h,j}a_h+b^{out}_j\\
    y_j&=f_2(\beta_j)= \mathrm{softmax}(\beta_j)=\frac{e^{\beta_j}}{\sum\limits_{l=1}^{K}e^{\beta_l}}. \\
    \end{array}
\end{equation}

For the goal of the ANN to classify asteroids using spectral data, the output $y_j$ of this ANN should be close to that of the known labeled class. If the weights and biases are properly fitted, the predicted values of the ANN should be consistent with the labeled values of the input data. Otherwise, these weights and biases involved in the hidden layer and the output layer need to be adjusted. In our case, there are $(N_{in}+1)N_h+(N_h+1)K=7300$ parameters (weights and biases) that need to be optimized to achieve accurate prediction of the input data classes. The procedure to find the best values for these involved parameters is also called a 'learning' or 'training' procedure. 

\subsection{Learning procedure} \label{subsec:Learning procedure}

The structure of our ANN provided, the back-propagation algorithm is applied to optimize the weights and biases. The cross-entropy measure (see Eq.~(4)) is used as the cost function of our ANN \citep{2005A}, which measures the difference between the predicted values and the labeled values in the training data. Given the labels of input data, the smaller value of loss function means a better performance of this network. If there are $N_{sample}$ training samples included in a training set, the loss function can be written as: 
\begin{equation}
    H(P,Q)=-\sum\limits_{n=1}^{N_{sample}}\sum\limits_{i=1}^{K}p_{(i,n)} \log(q_{(i,n)}) \ ,
\end{equation}
where vectors $P=([p_{(i,n)}], i=1,\ldots,K; n=1,\ldots,N_{sample}$) and $Q=[q_{(i,n)}]$ are the labeled and predicted values, respectively. The subscript $(i,n)$ refer to class $i$ and sample  $n$.

The learning procedure is actually optimization of the ANN where the weights and biases are adjusted in order to minimize the loss function. The optimization method we applied here is the mini-batch gradient descent (MBGD) algorithm \citep{Goodfellow-et-al-2016}. It is one variant of the stochastic gradient descents (SGD), and has both the merits of the SGD and the batch gradient descent (BGD). The MBGD involved a small part of samples picked randomly from the whole training data set, called a batch of samples. With a batch of samples,  the weights and biases are iteratively corrected until the cross-entropy reaches a minimum. The most-used approach in the optimization of the ANN is to move along the gradient direction of the loss function. Let a vector $z = [w,b]^T$ represent the parameters with $w$ as the weights and $b$ as the biases. The parameter vector is iteratively updated along to the negative gradient of the loss function with Eq.~(5). 
\begin{equation}
 \begin{array}{cl}
    \left[\begin{array}{c}
    w_{k+1}\\
    b_{k+1}\\
    \end{array}\right]=&
    \left[\begin{array}{c}
    w_{k}\\
    b_{k}\\
    \end{array}\right]-\alpha \left[\begin{array}{c}
    g1_{k}\\
    g2_{k}\\
    \end{array}\right]\\
    \left[\begin{array}{c}
    g1_{k}\\
    g2_{k}\\
    \end{array}\right]=&
     \left[\begin{array}{c}
    \frac{\partial H(P,Q_k)}{\partial w}\\
    \frac{\partial H(P,Q_k)}{\partial b}\\ 
    \end{array}\right]\\
 \end{array}
\end{equation}

From Eq.~(5), new step of update to the parameters $z_{k+1}$ is related to the previous values of parameters $z_{k}$ and the directions of gradient decrease of the loss function $g1,g2$. The quantity $\alpha$ ($> 0$), called the learning  rate, is usually set considering an efficient convergence of the training procedure.

Then, a new batch of samples is picked up from the rest of the training data set, and the parameters are continuously updated until each sample in the training data set is used. In practice, the entire training data set is randomly split into some batches according to the number of samples in a batch. The round when all batches are used in training the ANN is called a training epoch. Continuously, the training data set is split randomly and the next epoch of training of the ANN is done, until some conditions are satisfied, for examples, the variation of the loss function is smaller than a threshold or the number of epoch reaches the maximum limit. Based on our experience in training the ANN, the MBGD has a stable convergence speed, and is more efficient than the BGD.

\section{Data preparation} \label{sec:data}

Here we build an ANN tool to classify future spectral data of asteroids from the CSST and the new observed spectra from a ground-based telescope. The future CSST's spectral data of asteroids could be obtained by one of 12 spectroscopic units of the survey module (see Fig.~\ref{fig:ccd}, four for each of GU (0.255 $\sim$  0.4 \textmu{}m), GV (0.4 $\sim$ 0.6\textmu{}m), and GI (0.6 $\sim$ 1.0 \textmu{}m). Each spectroscopic unit consists of two gratings of opposite dispersion directions, one band-passing filter, and a CCD detector. The  +1 and -1 order spectrum of two gratings could give a stellar spectral image of 3.4 to 4.1~mm in length, corresponding to the equivalent spectral image a $300 \sim 400$ pixels in a $9K$ CCD image assuming a pixel size of 10~\textmu{}m. Spectra of asteroids could  cover a wavelength range of 0.255 to 1.0~\textmu{}m if the target passes the GU, GV, and GI spectroscopic units. Considering the wavelength range of CSST spectral data (0.255 to 1.0~\textmu{}m) and our new derived spectral data (0.40 to 0.83~\textmu{}m), the SMASS II data and the classification labels from the Bus-Binzel taxonomy system are chosen to construct our ANN tool.

To train and test the ANN of asteroid classification, we used the spectral data from the database of the Small Main-Belt Asteroid Spectroscopic Survey Phase II \citep[SMASS II]{2002Icar..158..146B,2002Icar..158..106B} and the Bus-Binzel taxonomy system \citep{2002Icar..158..146B}.  

\subsection{Sample selection}

The SMASS data are the first spectral data of asteroids obtained with the  spectroscopic instrument using a CCD detection, so more features than the broad-band spectrophotometric colors are detected. The SMASS II database contains spectral data of 1447 asteroids covering a wavelength range of 0.435--0.925 \textmu{}m. Each spectrum is composed of 182 data points with a sampling interval of 2.5 nm in wavelength.

To inherit from the Tholen taxonomy, \cite{2002Icar..158..146B} constructed a feature-based asteroid taxonomy with a PCA method based on the SMASS II data and information. It is called the Bus-Binzel or SMASS II taxonomic system, in which 26 classes (A, B, C,  Cb, Cg, Cgh, Ch, D, K, L, Ld, O, Q, R, S, Sa, Sk, Sl, Sq, Sr, T, V, X, Xc, Xk and Xe) are identified from the spectra of 1447 asteroids. 

The spectral data of SMASS II we used here is downloaded from a website\footnote{\url{https://sbnapps.psi.edu/ferret/}}.  Among the 1447 taxonomic asteroids in the SMASS II system,  1139 asteroids belong to C, S, and X-complexes, each of which are further divided into subclasses according to features in spectral data. The rest are identified as the special classes: A, B, D, K, L, O, Q, R, T, V.  The occurrence of subclasses in the Bus-Binzel system reflect the connection or transition to its closely neighboring classes \citep{2002Icar..158..146B}.
We consider that it is also important to understand the spectral variations due to heterogeneous surfaces and due to different observational geometry and epoch.

As the first stage of the analysis to the spectral data using CSST and ground-based telescopes, 10 classes (A, B, C, D, K, L, S, T, V, and X-type) are included in our ANN to have enough samples in each class. The classification to subclasses of asteroids will be considered in future work. In all, 834 samples are picked from the SMASS II database. The detailed numbers of samples for each class are listed in Table~\ref{tab:samples}. For convenience, we attributed integer numbers $0,\ldots,9$ to the classes we use here (A, B, C, D, K, L, T, V, X, and S-type).

\begin{table}
\caption{Number of samples for each taxonomic type in our data.}
\label{tab:samples}
\centering  
\begin{tabular}{c c c c c c c c c c c}    
\hline\hline       
Class & A & B & C & D & K & L & S & T & V & X  \\ 
\hline
$\#$ & 16 & 60 & 142 & 9 & 31 & 34 & 382 & 14 & 35 & 111 \\
\hline                  
\end{tabular}
\end{table}

\subsection{Resampling the SMASS II spectra}

The future slitless spectral data  of the CSST will cover a wavelength range of 0.255 to 1.0 \textmu{}m with a lower limitation of the resolution of 200 \citep{zhanhu}.  Before the construction of  the ANN for the CSST spectra, the training data, meaning the SMASS II spectra, need to be re-sampled to match the CSST observations. Considering the shortest wavelength of $\lambda_0=0.435$ \textmu{}m and the longest wavelength of 0.925 \textmu{}m, the re-sampled data of SMASS II are composed of 151 data points(Resampling is completed according to expression (1)).

\begin{figure*}[ht!]
\centering
\includegraphics[width=1\columnwidth]{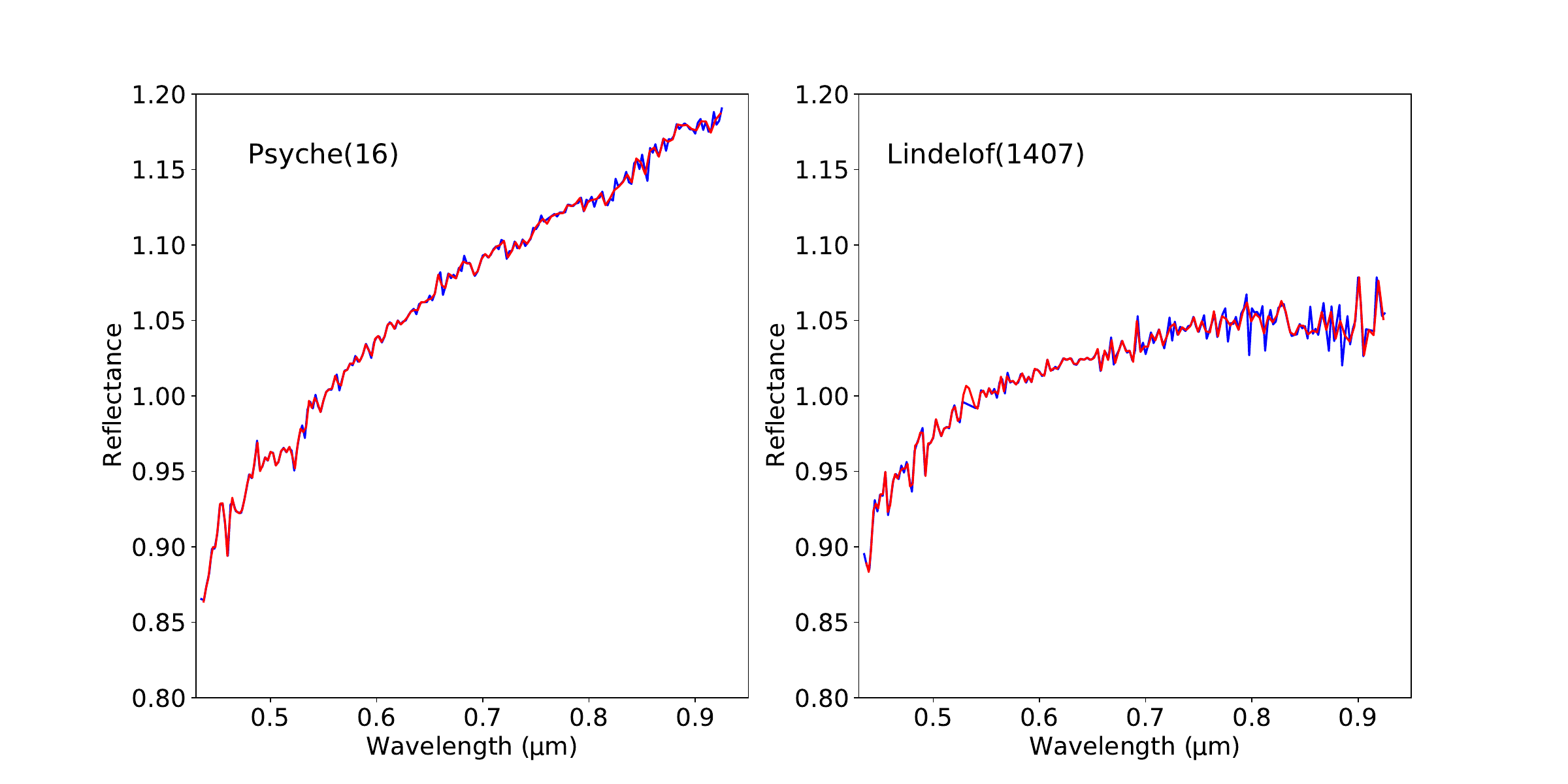}
\caption{The blue line is the original data of SMASS II, and the red line is the re-sampled data.}
\label{fig:or}
\end{figure*}

In the actually observed spectral data, as the wavelength increases, the number of data points will become more and more sparse. Using expression (1) to resample will be closer to the actual observation data. The  reflectance value corresponding to each re-sampled wavelength is obtained by cubic spline fitting to the spectral data. Considering the wavelength interval of original spectra of SMASS II, we set 0.02 \textmu{}m as the bin interval of cubic spline fitting, so at least 10 data points are involved in fitting each piece-wise spline function. Most re-sampled spectra are very close to the original ones, only some very noisy data are smoothed in the spline fitting procedure (see Fig.~\ref{fig:or}).  The re-sampling procedure gives us 834 spectra, each having 151 features/channels.  

\subsection{Cloning of spectra}

From Table \ref{tab:samples}, it is easy to notice that the sample counts of different classes are not balanced, for example, the D-type has only 9 samples whereas the S-type has the largest number of samples, 382. If such data is used to train the ANN,  the neural network could optimize and learn the features of the classes with large numbers of samples only. To balance the samples in different classes,  we added some synthetic spectra into those classes with few samples. The synthetic spectra of a class are generated by a clone method based on all real data in that class.

\begin{figure*}[ht!]
\centering
\includegraphics[width=1\columnwidth]{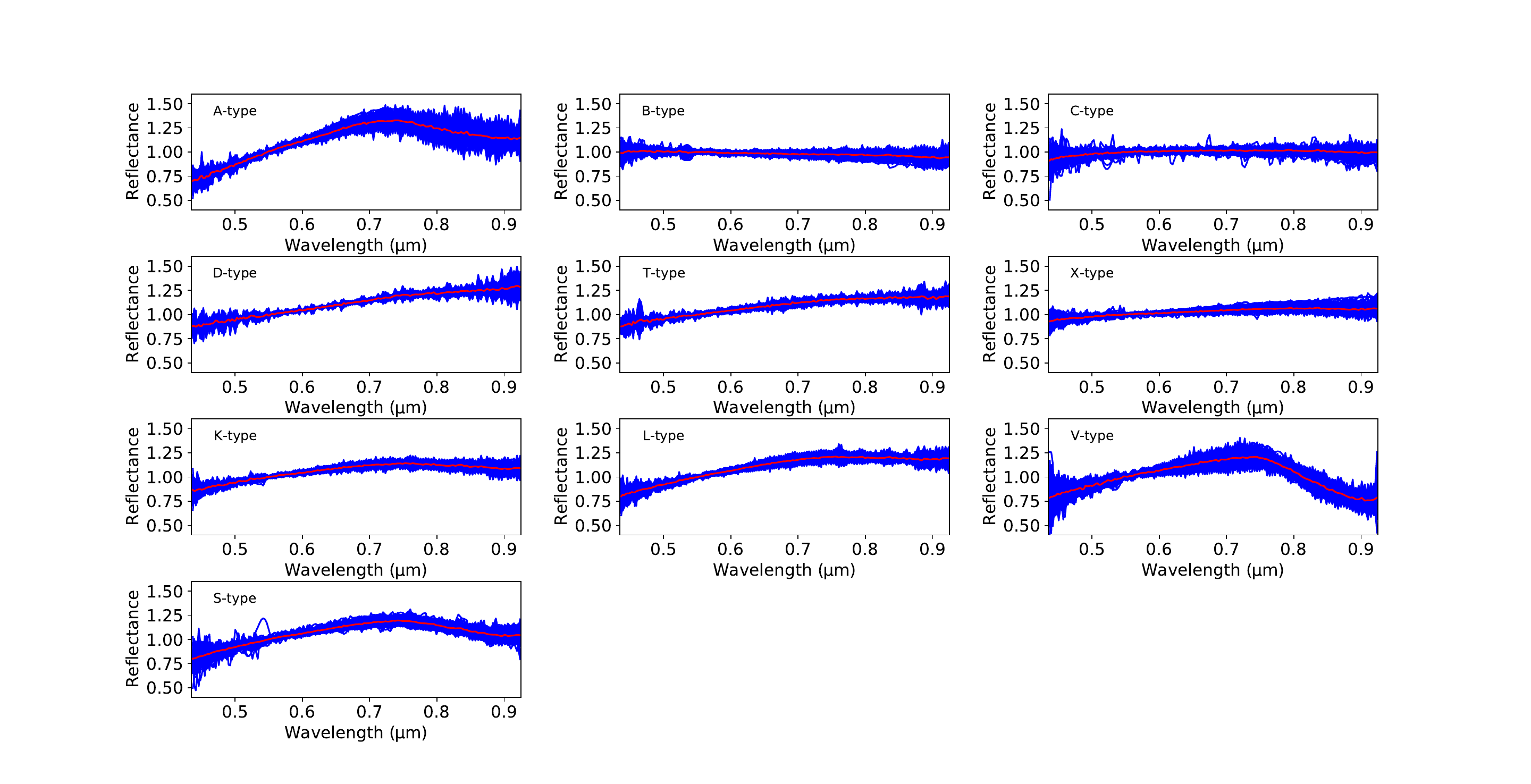}
\caption{Input data/spectra for the ANN. Blue color shows the real and cloned spectra, and the red line represents the average spectrum for each class.
\label{fig:train}}
\end{figure*}

The spectra of asteroids of the same class show similar shapes with some variation.  To obtain the synthetic spectra, first, the representative spectrum of each class is found by averaging all re-sampled spectra in the class (the red line in Fig.~\ref{fig:train}). In detail, we calculated the mean reflectance at wavelengths $\lambda_i$ (i = 1, …, 151) and the corresponding standard deviations $\sigma^i_k$ (with $k$ as class index, $k=0,\ldots,9$) over all re-sampled spectra in the class $k$. Then, a clone spectrum of class  $k$ is simulated by adding Gaussian noise with standard deviation of $\sigma^i_k$ into the representative spectrum of the $k$ class. Repeatedly, cloned spectra are generated until the total number of spectra in each class reaches 400. The cloned and real spectra for each classes are shown in  Fig.~\ref{fig:train}. The $10\times400$ spectra form the dataset used in training, validating, and testing our ANN in the next step. During the construction of the ANN, the dataset is divided into training, validation, and test data sets. 

\section{Construction of the ANN } \label{sec:con}

The construction of the ANN is a procedure to find an optimal mapping relationship between the input data and the labels. In detail, for an ANN of three layers (the input layer of $N_{in}=151$ neurons, one hidden layer of $N_h=45$ neurons, and the output layer of $K=10$ neurons), there  are $(N_{in}+K+1)\times N_h+K=7300$ parameters to be optimized.

To use all spectra in every optimization round would result in a time-consuming and heavy procedure. Therefore we applied the MBGD (Mini Batch Gradient Decrease) algorithm in optimization with multiple parameters. With the algorithm, some additional parameters are introduced: batch size, number of epochs, learning rate, and the dropout layer ratio. Batch size is the number of samples in each batch as the training data set is randomly divided into smaller batches. One training round means training the ANN with one batch. When all batches are used to update the weights and the biases, one epoch of training is done on the ANN.  So, the quantity of epochs determine how many times we split randomly the training data set. When setting this value, one needs to consider the computation time and the convergence of the training procedure. Learning rate controls the step size when updating the parameters in each iteration steps. Too large a learning rate may cause oscillating values for the parameters.  Using the ReLU function as the transfer function between the input layer and the hidden layer,  a small learning rate is preferred.

In the beginning of training, the parameter space grid (batch size, epochs, learning rate) is scanned to find good values for the three parameters. Each combination of the three parameters is tested by checking the efficiency of the training performance. Finally, the batch size of 64, the epoch number of 1000, and the learning rate of 0.001 were chosen. All the ANN parameter values are listed in Table \ref{tab:model}.

\begin{table}
\caption{ANN parameters.}
\label{tab:model}
\centering  
\begin{tabular}{l l}    
\hline\hline       
Property & Value  \\ 
\hline
Neurons on input layer  & 151\\
Neurons on hidden layer  & 45 \\
Hidden layer activation function & ReLu\\
Neurons on Output layer & 10 \\
Output layer activation function & Softmax \\
Training algorithm & MBGD \\
Batch size & 64 \\
Learning rate & 0.001 \\
Epochs & 1000 \\
\hline                  
\end{tabular}
\end{table}

\subsection{ANN training } \label{subsec:ann training}

We have an input dataset of 4000 spectra to train and test the ANN, and a ratio of 4:1 is set to divide randomly the 4000 spectra into the training and test datasets. The training procedure of the ANN starts with initializing the weights and biases by random values, and then those weights and biases are optimized iteratively by the training procedure with the following steps: (1) separate the input dataset into training and test data set with the ratio 4:1; (2) split randomly the training data set into batches according to the batch size of 64; (3) correct the weights and the biases by the gradient descent method to minimize the loss function, in which the back-propagation method is used to implement the optimization; (4) repeat to update the weights and the biases with another new batch of data until all batches are used; that is one epoch of iteration to construct the ANN; (5) repeat (1)--(4) to reach the required number of epochs; (6) check and  test the ANN performance with the test dataset.

Additionally, we introduced a dropout layer in our ANN to avoid an over-fitting problem. The ratio of dropout layer is set as 0.2, which means that random 20 \% of neurons in the hidden layer are not connected to the neurons in the output layer in each round of training.

The optimization with gradient decrease (GD) family methods often finds a local minimum directly related to the initial values of the optimized parameters. To overcome this, five ANNs with different initial values for the parameters are constructed, and the final classification result to an input data is determined by voting among the five ANNs.

\subsection{ANN accuracy} \label{subsec:ANN accuracy}

Using 834 real spectra of the SMASS II dataset, the classification performance of the trained ANN is assessed using the classification accuracy, the count ratio of correctly predicted cases to the entire sample. When using the softmax function as the activation function of the output layer, the ANN outputs a probability distribution of the taxonomic classes. The final predicted label of a spectrum is the class with the maximum probability in the probability distribution. To make the prediction result more robust, the vote result of five predicted labels suggested by five ANNs, initialized with different random numbers,  is taken as the final predicted label for the input sample.

\begin{figure*}[ht!]
\centering
\includegraphics[width=0.8\columnwidth]{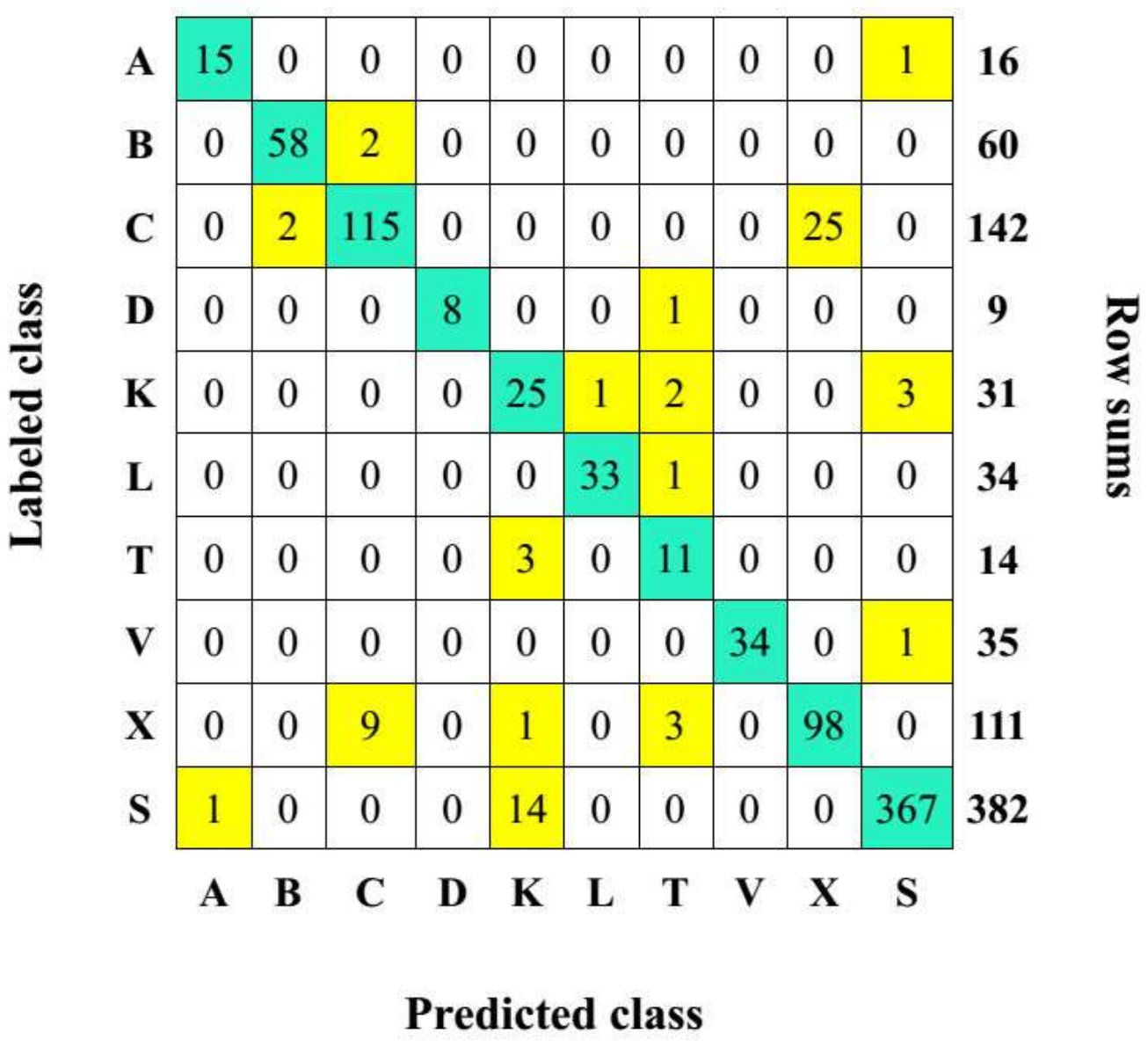}
\caption{The confusion matrix. Each diagonal element with green color indicate the number of samples having consistent labels with the Bus-Binzel taxonomy, given by the row index, while the elements in that row with yellow color show the number of samples mislabeled into a class corresponding to the column of the value.
\label{fig:matrix}}
\end{figure*}

We investigated the accuracy of the classification ANN by comparing the final predicted labels for the 834 chosen real spectra from SMASS II to their labels in the Bus-Binzel taxonomic system. Generally, 764 out of 834 asteroids are correctly classified, which gives an accuracy of 92 \%. Considering the accuracy of each class, the accuracies of A, B, L, and S-class are higher than 94 \%, whereas the C, D, T, X, and K-classes have slightly lower accuracies. The detailed values for the 10 classes are listed in Table \ref{tab:acc}. From the confusion matrix (see Fig.~\ref{fig:matrix}), we can see where the problematic cases are.  The diagonal elements of the confusion matrix are the number of samples with predicted labels consistent with that in the Bus-Binzel taxonomy,  and conflicting  labels are the off-diagonal elements of the matrix.  It is easy to note that the top conflicting labels are between the C and X-class, and between the S and K-class. For example,  25 C-class asteroids are predicted to be X-class, and 14 S-class asteroids were labeled as K-class with our ANN. Among the 25 C-class samples in the Bus-Binzel system labeled as X-class by our ANN, we found 9 to show a gentle convex behaviour between 0.6--0.8 \textmu{}m, indicating that they might belong to the Xc-class or the transition between the C and X-classes. The 14 S-class samples in the Bus-Binzel system are predicted as K-class by our ANN probably because the classes show very similar spectral shapes.

\begin{table}
\caption{The classification accuracy of our ANN tool.}
\label{tab:acc}
\centering  
\begin{tabular}{c c c c c c c c c c c c c}    
\hline\hline       
Class & A & B & C & D & K & L & S & T & V & X  & Averaged\\ 
\hline
Accuracy & 0.94 & 0.97 & 0.81 & 0.89 & 0.81 & 0.97 & 0.96 & 0.79 & 0.97 & 0.88 &0.92\\
\hline                  
\end{tabular}
\end{table}

\subsection{Test with the Small Solar System Objects Spectroscopic Survey data} \label{subsec:tables}
As additional test of our ANN, we selected spectra of 415 asteroids from  the Small Solar System Objects Spectroscopic Survey (S3OS2) \citep[for short, S3OS2]{2007PDSS...51.....L}, which are obtained with the 1.52-m telescope of the European Southern Observatory during the period of November 1996 and September 2001. The wavelength of the S3OS2 data range 0.49-0.92~\textmu{}m. \cite{2007PDSS...51.....L} classified the S3OS2 data refering to the Bus taxonomy system, the analysis results denoted here as S3OS2(B). Among the 415 samples picked up from the S3OS2 dataset, 80 are of S-type, 205 of B, C, or X-type, and 130 of A, D, K, L, T, or V-type.

With a similar procedure, those test data are firstly re-sampled according to the format of the input data of our ANN. Then, they are analyzed with our ANN. If comparing our classification results to those in S3OS2(B), 362 out of 415 samples have the same taxonomy labels. The ratio of the matched samples to the whole data is $\sim$ 87 \%. In  detail, the accuracies for S-type, the B/C/X-type group, and the A/D/K/L/T/V-type group are 0.89, 0.87, and 0.86, respectively.

\section{Application} \label{sec:app}
As the first application of our ANN tool, our observed spectral data obtained by our group with the 2.16-m telescope in the Xinglong site of National Astronomical Observatory (observatory Code 327) in years 2006--2007 are analyzed. The aims of the spectroscopic observations are to figure out the diversity of spectra among primitive asteroids, and the spectral variations of individual asteroid at different rotational phase.  So, we have chosen targets which are primitive asteroids (i.e., C-complex ) and could be observed well with the 2.16-m telescope of National Astronomical Observatory of China in the Xinlong site during the allowed time period (with a small zenith distance at asteroid's meridian passage). With our ANN tools, those spectral data mentioned above are analysed. In what follows, detailed information is given on the spectroscopic observations, data reduction, and the analysis results of classification.

\subsection{Observations and data processing} \label{subsec:Observations}

The five nights of spectroscopic observations were carried out in three nights in 2006 (December 26th, 27th, and 28th) and in two nights in 2007 (November 17th and 19th). The spectral data of the selected asteroids were obtained by an OMR cardigan spectrometer and a PI $1340\!\times\!400$ CCD detector. A grating of 300 lines/mm of the OMR and a long slit of 2.5'' in width  gives a resolution of $R = 200$. The orientation of the long slit of the OMR instrument is along the South-North direction.  At each night of observations, at least two solar analog stars (selected from HR4030, HR3538, HR3951, HD28099, HD191854, and HR996) were observed one or two times. For each observation of asteroids and solar-like stars, one spectrum of a HeAr lamp installed in the OMR system was followed for the wavelength calibration. In order to eliminate the effect of atmospheric extinction, we arranged the spectroscopic observation of asteroids at the smallest possible zenith distance.

Because of the motion of an asteroid relative to the stars in the same field of view, it is possible for the asteroid to closely pass by a star, and that could result in stellar light entering the slit, which would lead to contamination of the spectral data of the asteroid.  We checked the observation images of asteroids, and contaminated data are picked up and rejected. Also, part of the spectral data on Dec. 28th, 2006 are not used due to the abnormal operation of the OMR. Finally, 64 spectra of 42 asteroids are involved in the analysis. In Table \ref{tab:observation}, the detailed observational information, including the time of observation, exposure time, airmass, and the solar analog stars used, is listed for each spectrum. 

\begin{table}
\caption{Observation information for 42 asteroids}
\label{tab:observation}
\centering  
\resizebox{\textwidth}{!}{
\begin{tabular}{c c c c c | c c c c c }    
\hline\hline       
Asteroid & Observation & Exposure & Airmass & Solar analog & Asteroid & Observation & Exposure & Airmass & Solar analog\\ 
 & time(JD) & time(s) &  & stars & & time(JD) & time(s) & & stars \\

\hline
326 & 2006-12-26.546 & 1800 & 1.262 & HR4030  &34 & 2007-11-17.681 & 1800 & 1.186 & HD28099\\
 & 2006-12-27.485 & 2400 & 1.103 & HR3538  && 2007-11-17.702 & 1800 & 1.241 & HD28099\\
407 & 2006-12-26.736 & 600 & 1.271 & HR4030 &35 & 2007-11-17.470 & 2400 & 1.402 & HD28099 \\

468 & 2006-12-26.818 & 3600 & 1.143 & HR4030 & & 2007-11-17.497 & 2400 & 1.443 & HD28099  \\

469 & 2006-12-26.643 & 1800 & 1.213 & HR4030 &95 & 2007-11-17.783 & 1200 & 1.160 & HD28099\\

& 2006-12-26.746 & 1800 & 2.128 & HR4030 &  & 2007-11-19.730 & 900 & 1.078 & HD191854  \\
 
& 2006-12-26.668 & 1800 & 1.329 & HR4030 &105 & 2007-11-17.531 & 3600 & 1.441 & HD28099 \\

 & 2006-12-27.444 & 3000 & 1.072 & HR3538  &159 & 2007-11-17.806 & 1800 & 1.126 & HD28099 \\
  
 & 2006-12-27.516 & 3000 & 1.007 & HR3538  &381 & 2007-11-17.837 & 3000 & 1.581 & HD191854\\

 & 2006-12-27.615 & 3000 & 1.149 & HR3538  && 2007-11-17.872 & 3000 & 2.003 & HD191854\\

 & 2006-12-27.729 & 3000 & 2.010 & HR3538 &410 & 2007-11-17.751 & 2400 & 1.128 & HD191854\\
  
 & 2006-12-28.486 & 1800 & 1.020 & HR4030 &444 & 2007-11-17.726 & 1200 & 1.224 & HD191854\\
 
 & 2006-12-28.587 & 1800 & 1.062 & HR4030 &567 & 2007-11-17.577 & 3600 & 1.162 & HD191854\\
 
 494 & 2006-12-26.692 & 1200 & 1.109 & HR4030  & & 2007-11-17.619 & 3600 & 1.195 & HD191854\\

 514 & 2006-12-26.879 & 3000 & 1.372 & HR4030 &98 & 2007-11-19.508 & 1200 & 1.044 & HD191854 \\
 
 613 & 2006-12-26.597 & 1800 & 1.272 & HR4030 &120 & 2007-11-19.685 & 900 & 1.061 & HD191854\\
  
 & 2006-12-26.619 & 1800 & 1.397 & HR4030 &146 & 2007-11-19.881 & 1800 & 1.370 & HD191854\\
 
 663 & 2006-12-26.498 & 3600 & 1.412 & HR4030 &168 & 2007-11-19.651 & 900 & 1.117 & HD191854 \\
 
 696 & 2006-12-26.916 & 900 & 1.647 & HR4030 & 445 & 2007-11-19.435 & 1800 & 1.046 & HR996 \\
 
 & 2006-12-27.842 & 2400 & 1.254 & HR3538 &362 & 2007-11-19.717 & 900 & 1.048 & HR996\\
 
 1004 & 2006-12-26.773 & 3600 & 1.149 & HR4030 &508 & 2007-11-19.763 & 1800 & 1.160 & HR996\\
 
 & 2006-12-28.699 & 2400 & 1.108 & HR4030  &566 & 2007-11-19.699 & 900 & 1.093 & HR996\\
 
 414 & 2006-12-27.692 & 3000 & 1.265 & HR3951 &583 & 2007-11-19.463 & 3600 & 1.133 & HR996\\
  
 772 & 2006-12-27.555 & 2400 & 1.115 & HR3538 &602 & 2007-11-19.525 & 600 & 1.052 & HR996 \\
  
 & 2006-12-27.584 & 2400 & 1.170 & HR3538 &733 & 2007-11-19.834 & 3600 & 1.109 & HD28099  \\
 
 780 & 2006-12-27.767 & 2400 & 2.636 & HR3538  &786 & 2007-11-19.647 & 3000 & 1.247 & HD28099\\
 
 & 2006-12-27.799 & 2400 & 2.612 & HR3538 &850 & 2007-11-19.788 & 3600 & 1.441 & HD28099\\
 
 1911 & 2006-12-27.883 & 3600 & 1.432 & HR3538  &907 & 2007-11-19.719 & 1200 & 1.042 & HD28099\\
 
680 & 2006-12-28.515 & 3000 & 1.012 & HR4030 & 977 & 2007-11-19.556 & 3600 & 1.305 & HD28099 \\

& 2006-12-28.551 & 3000 & 1.032 & HR4030 &&&&& \\
199 & 2006-12-28.649 & 1200 & 1.028 & HR4030 &&&&& \\
 & 2006-12-28.664 & 1200 & 1.032 & HR4030 &&&&&\\
1303 & 2006-12-28.437 & 1800 & 1.218 & HR4030 &&&&& \\
& 2006-12-28.460 & 1800 & 1.190 & HR4030 &&&&&\\
1512 & 2006-12-28.604 & 3600 & 1.017 & HR4030  &&&&&\\

\hline                  
\end{tabular}
}
\end{table}

The data reduction of the spectroscopic observations of asteroids were done according to the standard procedure \citep{2002aste.book..169B} with the IRAF software. Firstly, the systematic errors of the spectroscopic image, i.e., bias, dark current, and flat-field effects were corrected, then cosmic rays in the image were identified by a threshold of 4 times the standard deviation of the sky background and removed. Secondly, one-dimensional spectra of asteroids, solar-like stars, and HeAr lamp were extracted by an optimal aperture and a fitted background. The wavelength calibration of a target's spectrum was done with the aid of the spectrum of the HeAr lamp after the target. As for the atmospheric extinction effect, we used an average extinction curve of the Xinglong site (Code 327).

At each observational night, multiple bias frames were obtained at the beginning and end of observation of that night, and were combined into a synthetic bias image by averaging, called the best bias image.  All other spectroscopic images, i.e., those of the targets, flats, and HeAr lamp, were subtracted with the best bias image to correct the bias effect of the CCD detector. Similarly, a combined flat image was generated from multiple observed flat images which actually are spectroscopic images of an arc lamp in the telescope.  The illumination and response effect on the combination flat was simulated and then removed from the combination flat image.  The final flat image is a normalized combination flat image without  the illumination and response effect. The flat-field effect of the CCD detector on all spectroscopic images are corrected by dividing with the final flat image.

When extracting one-dimensional spectra of celestial objects, a trace procedure was applied. That is to trace the peak flux along the dispersion direction in the two-dimensional spectroscopic image  with an optimal width (or aperture), after which all flux within the aperture are summed along  the vertical dispersion direction to form the one-dimensional spectrum. The optimal aperture of the extraction of the one-dimensional spectrum is determined by balancing the maximum target flux and the minimum of background flux in the aperture.

For the wavelength calibration of a target spectrum, we used a Legendre function with the order of three or five. The calibration is derived by fitting the pixel positions of the emission lines for the spectrum of the HeAr lamp following the target to their wavelength values. On average, the wavelength range of our spectra is between 0.40 and 0.83~\textmu{}m. 

To reduce the influence of extinction as much as possible,
we scheduled the observation epochs for each asteroid for the time when they are as high in zenith as possible. In practice, we applied an average extinction curve of the Xinglong site of National Astronomical Observatory: the correction of extinction for each source spectrum is done according to the airmass of the time of the observation.

To obtain the reflectance of an asteroid, we conducted spectroscopic observations for the selected solar analog stars in the same night of the asteroid observations. Both the spectra of the asteroids and the solar analog stars are normalized to unity at the wavelength of 0.55~\textmu{}m, then the reflectance of an asteroid is that of the normalized spectrum of the asteroid divided by the normalized spectrum of the selected solar analog star. Sometimes multiple spectra of single solar analog star are obtained during one night, the averaged spectrum of those is then used to derive the reflectance of the asteroids. If spectra of multiple solar analog stars are obtained, we prefer to use the spectrum of a relatively stable star, or the averaged spectrum of those stars with no obvious activity.

\begin{figure*}[ht!]
\centering
\includegraphics[width=1\columnwidth]{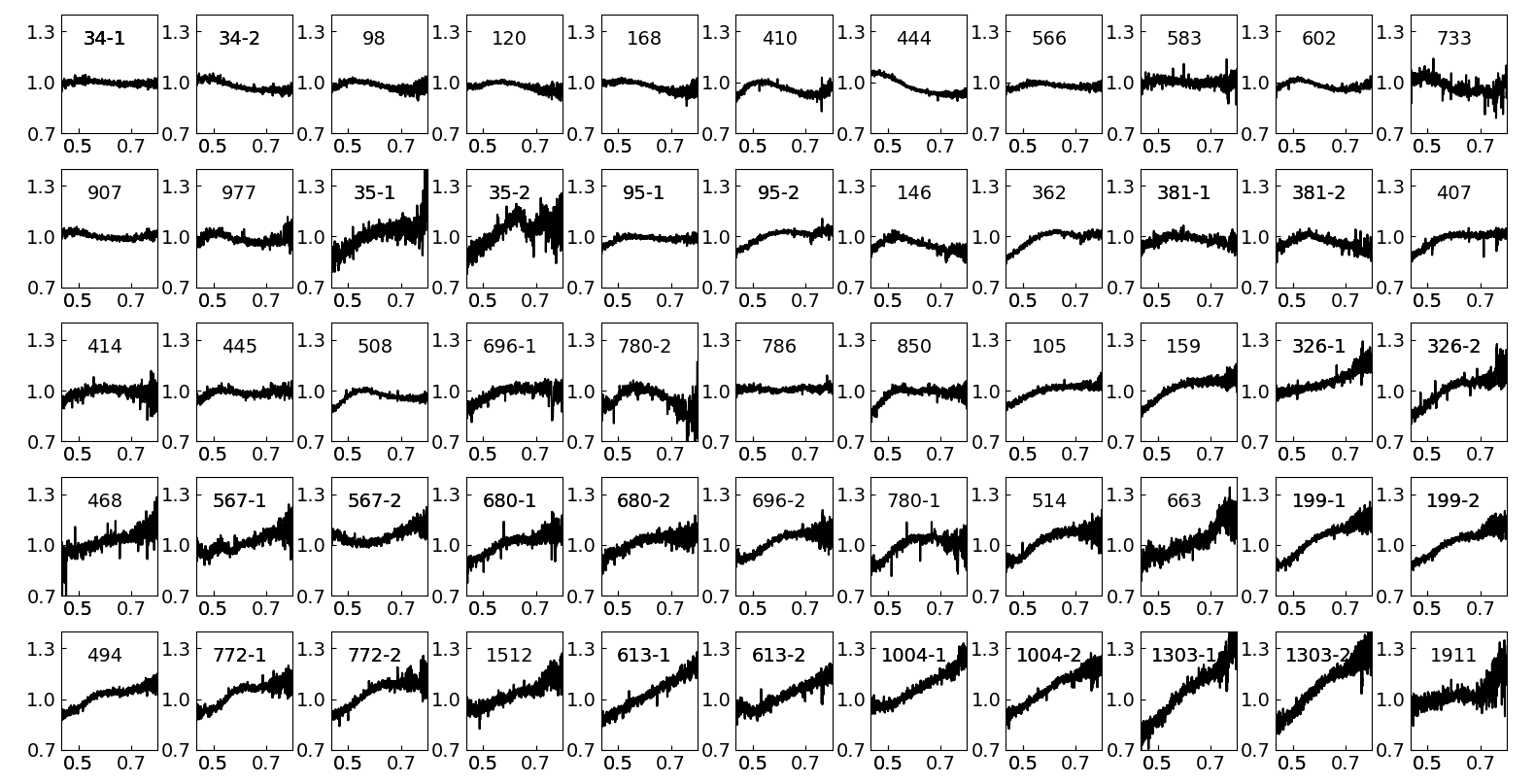}
\caption{55 reflectance spectra for 41 asteroids.
\label{fig:all}}
\end{figure*}

The sample of 55 reflectance spectra for 41 asteroids is shown in  Fig.~\ref{fig:all} and an example with 9 reflectance spectra of asteroid (469) Argentina is shown in Fig.~\ref{fig:469-types}.

\subsection{Taxonomy analysis with our ANN tool} \label{subsec:Taxonomy analysis}

According to the format of the input data of our ANN tool (151 channels ranging from 0.43 to 0.92~\textmu{}m  with an increase of the span of $(1+\frac{1}{R})^i$  $(i=1,\ldots,151)$, the 64 spectra of 42 asteroids are re-sampled using the cubic spline algorithm. Because the wavelength range of our data is around between 0.40 and 0.83~\textmu{}m, the shortage of data beyond 0.83~\textmu{}m is complemented by linearly extrapolated values.

\begin{table}
\caption{Classification results for 42 observed asteroids.}
\label{tab:result}
\centering  
\resizebox{\textwidth}{!}{
\begin{tabular}{l l l l l| l l l l l}    
\hline\hline   
\renewcommand{\thefootnote}{\fnsymbol{footnote}}
Targets & Observation & Type & S'\footnote[1]{Wavelength range used to compute spectral gradient S’ \citep{2002AJ....123.1039J,2018Icar..302...10L}: 0.55 - 0.86~\textmu{}m.} & Existing classification \footnote[2]{References for taxonomy: T84, \cite{1984atca.book.....T}; B02, \cite{2002Icar..158..146B}}. & Targets & Observation & Type & S' & Existing classification \\ 
 & date & & (\%/1000\AA) & results &&date && (\%/1000\AA) & results \\

\hline
34 & 2007-11-17.681 & B &-0.185& C(T84);Ch(B02) & 468 & 2006-12-26.818 & X&3.693&\\
 & 2007-11-17.702 &B &-2.344&&567 & 2007-11-17.577 & X&2.943& \\
98 & 2007-11-19.508 & B &-2.099& CG(T84);Ch(B02)& & 2007-11-17.619 & X&3.996 &\\
120 & 2007-11-19.685 & B &-2.784& C(T84);C(B02)&680 & 2006-12-28.515 &  X&3.221&\\
168 & 2007-11-19.651 & B &-2.636& C(T84);Ch(B02)& & 2006-12-28.551 & X&2.876&\\
 410 & 2007-11-17.751 & B &-2.321& C(T84);Ch(B02)&696-2 &2006-12-27.842 & X &2.755&\\
444 & 2007-11-17.726 & B&-3.456& C(T84);C(B02)&780-1 & 2006-12-27.767 &X&1.194&\\
566 & 2007-11-19.699 & B &-1.433& C(T84)&514 & 2006-12-26.879 & K&3.199& EMPC(T84) \\

583 & 2007-11-19.463 & B &-0.144& C(T84)&663 & 2006-12-26.498 & S &4.294& EMP(T84)\\
602 & 2007-11-19.525 & B&-1.162& C(T84)&199 & 2006-12-28.649 & T&5.913& X(B02)\\
733 & 2007-11-19.834 & B&-1.966& CF(T84)&& 2006-12-28.664 &T&4.689&\\
907 & 2007-11-19.719 & B &0.110& C(T84);Xk(B02)& 469 & 2006-12-26.643 &T&6.257& EMP(T84)\\
977 & 2007-11-19.556 & B&-0.510&&& 2006-12-26.668 & T   & 4.677&\\
35 & 2007-11-17.470 & C &1.638& C(T84);C(B02) && 2006-12-27.444 & T &5.681&\\
 & 2007-11-17.497 & C  &1.601& & & 2006-12-27.516 &  T& 5.054&\\
95 & 2007-11-17.783 & C&-0.284& C(T84);Ch(B02) & & 2006-12-27.615 &  T&5.525&\\
 & 2007-11-19.730 & C & 1.308&& & 2006-12-27.729 &  T &5.834& \\
 146 & 2007-11-19.881 & C&-5.091 & C(T84);Ch(B02)&& 2006-12-28.486 & T &6.078&\\
 362 & 2007-11-19.717 & C  & 0.064& & & 2006-12-28.587 & T  &5.808& \\
381 & 2007-11-17.837 & C & -1.676& C(T84);Cb(B02) & 494 & 2006-12-26.692 & T&3.939 & Ch(B02)\\
& 2007-11-17.872 & C&-2.991&&772 & 2006-12-27.555 & T&4.699& C(T84)\\
407 & 2006-12-26.736 & C&0.253 & C(T84)&& 2006-12-27.584 & T&5.109&\\
414 & 2006-12-27.692 & C&-2.099 & C(T84);Cg(B02)&1512 & 2006-12-28.604 & T&5.491& P(T84)\\
445 & 2007-11-19.435 & C &-0.120& &613 & 2006-12-26.597 &D&6.848& P(T84)\\
469 & 2006-12-26.746 & C&-2.029& && 2006-12-26.619 &D&4.889&\\
508 & 2007-11-19.763 & C &-2.462& C(T84)& 1004 & 2006-12-26.773 & D &8.776& \\
696-1 & 2006-12-26.916 & C&-0.680&&& 2006-12-28.699 &D&7.642&\\
780-2 & 2006-12-27.799 & C&-0.827&&1303 & 2006-12-28.437 & D&7.625&\\
786 & 2007-11-19.647 &C&0.965& C(T84)&& 2006-12-28.460 & D&9.918&\\

850 & 2007-11-19.788 & C&-0.275&& 1911 & 2006-12-27.883 &  D & 6.554&P(T84) \\
105 & 2007-11-17.531 & X&1.515& C(T84);Ch(B02)&\\
159 & 2007-11-17.806 & X &3.053& C(T84);Ch(B02)&\\
326 & 2006-12-26.546 & X&4.541& C(T84)& &  & & & \\
& 2006-12-27.485 &  X & 4.533&&& &  & &\\
\hline                  
\end{tabular}
}
\end{table}

These 64 re-sampled spectra are analysed with our ANN tool. The classification results for the 64 spectra of the 42 observed asteroid are listed in Table \ref{tab:result}. Statistically, among those 64 spectra, 17 are predicted as C-class, 13 as B-class, 11 as X-class, 14 as T-class, 7 as D-class, and S and K-classes have one sample each. 

Among the 42 observed asteroids, 27 have only a single spectrum, 14 have two spectra, and (469) Argentina has nine spectra.  The nine spectra of (469) Argentina obtained in three adjacent nights in 2006, obtained T-class, except one of a label of C-class (see Fig.~\ref{fig:469-types}). For the case of two spectra, spectral data of ten asteroids (34, 35, 199, 381, 567, 613, 680, 772, 780, and 1303) were obtained in the same night, while four asteroids (95, 326, 696, and 1004) were obtained in two adjacent nights. The analysis results with our ANN tool show that asteroids with two spectra obtained consistent labels except for asteroids (696) and (780). These two asteroids obtained C and X-class spectra (see Fig.~\ref{fig:xc-types}).
 
\subsection{Comparison to existing taxonomic information} \label{subsec:Comparison}

We compared our results of observed targets to their existing classification (see Column 5 in  Table \ref{tab:result}). Among the 42 observed asteroids, 31 have taxonomic type in the Tholen system  \citep{1984atca.book.....T} and/or the Bus-Binzel system \citep{2002Icar..158..146B}.  For these 31 labeled asteroids, 24 are of C-type or sub-class of the C-complex in Tholen and/or Bus-Binzel system,  three are of E/M/P-type  and another three are of P-type, and one is X-type.

Asteroid (199), labeled as X-type in the Bus-Binzel system, is classified as T-type with our ANN tool based on its two spectra. Three P-type asteroids (613, 1512, and 1911) in the Tholen system are identified here as D-class (613 and 1911) and T-class (1512). For the three E/M/P-type asteroids in the Tholen system, (469) is identified as T-class, (514) as K-class, and (663) as S-class. Considering the large albedo of E/M/P-type asteroids, they usually have a moderate spectral slope in the visible band. The new labels for the three asteroids are reasonable.

The 24 asteroids labeled as C-type or sub-class of C in the Tholen system are classified as C, B, and X-class (in detail, 8 asteroids as C, 11 as B, 3 as X and 2 as T) by our ANN tool. 
Checking the data of the Bus-Binzel taxonomic system, we found that 16 of the 42 observed asteroids are included, and fourteen asteroids were labeled as sub-class in the C-complex (C, Ch, Cb, or Cg-type), and two asteroids (199 and 907) as X-complex.  For these fourteen C-complex asteroids,  six are classified as B-type, five as C-type, two as X-type, and one as T-type by our ANNs tool.  The asteroid (199) is classified  as T-type  and the asteroid (907) as B-type by our ANN tool. 

Generally,  asteroids of C-type in the Tholen system or C-complex in the Bus-Binzel system are classified as one of the C, B, X, and T-types with our ANN tool. We think this is due to the featureless spectra of the B, C, X, and T-types. Some features, i.e., the slope between 0.55 and 0.86~\textmu{}m, shallow absorption in UV,  absorption around  0.70~\textmu{}m, or maybe  absorption around 0.85~\textmu{}m (see Fig.~\ref{fig:train}) are key points to distinguish the above four types. The reasons of confusion between the four types could be the low signal-to-noise ratio in the UV-band and the lack of data above 0.83~\textmu{}m of our data. Anyway, we could not rule out the spectral variations of an asteroid rising from heterogeneous composition over its surface and/or change of the observational geometry, e.g., solar phase angle.

\subsection{Spectral variation of asteroids at different rotational phases} \label{subsec:Spectral variation }

In our data, 14 asteroids have two spectra, and one asteroid has 9 spectra obtained in three adjacent nights. For the aim of classification of asteroids, we built five ANNs to determine the type for each spectra by voting. To check the final type of each spectrum for those asteroids with more than one spectrum, 13 asteroids received a consistent type, two asteroids have different labels (see Table \ref{tab:result}).  To understand the subtle spectral variation of an asteroid at different rotational phase, we choose one ANN from our five ANNs which gives the most similar result as compared to the voted result. Here, we compared the maximum probability output by the chosen ANN of each spectrum of asteroids with multiple spectra.

\begin{figure*}[ht!]
\centering
\includegraphics[width=1\columnwidth]{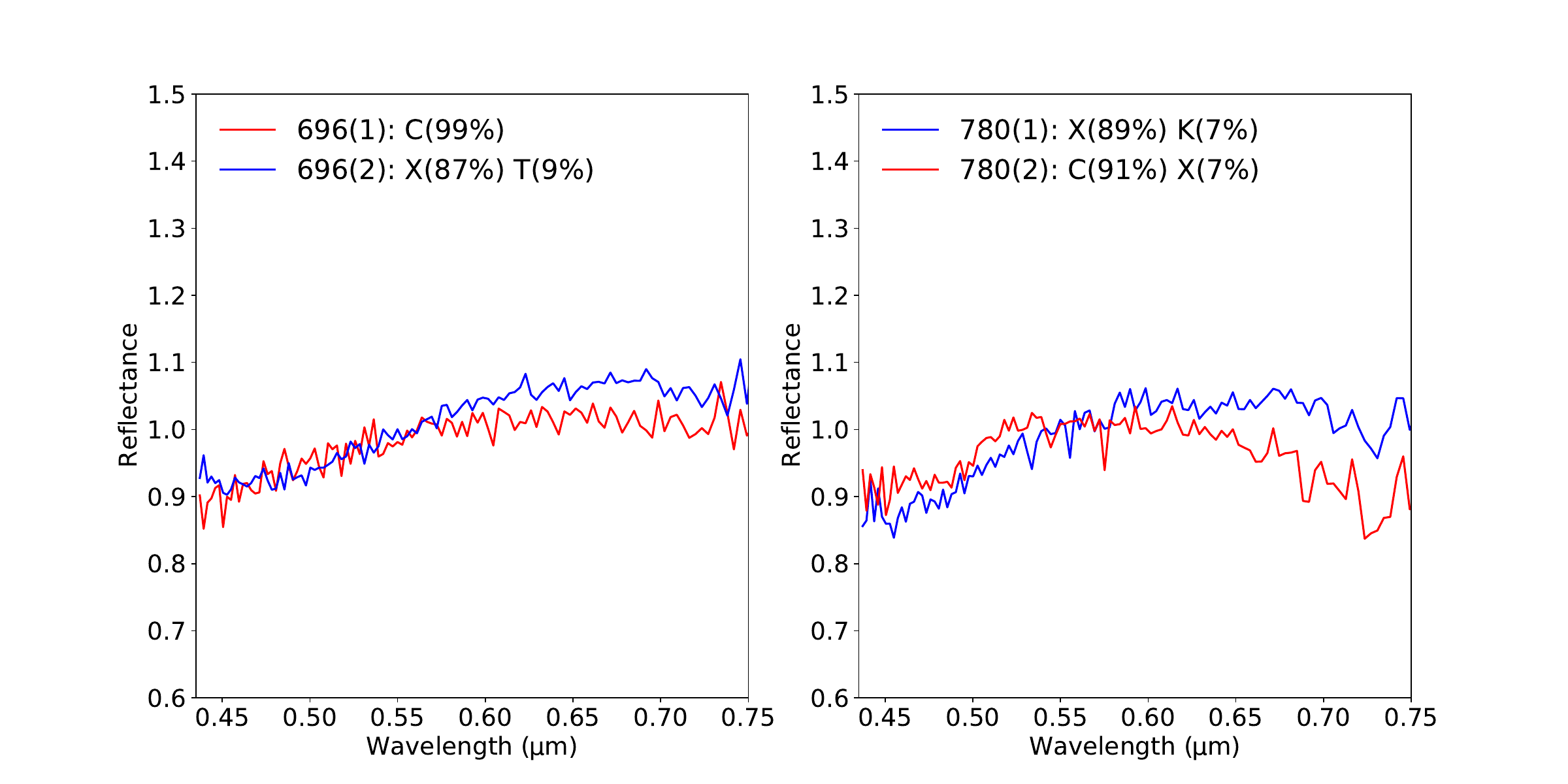}
\caption{Spectra of asteroids (696) Leonora and (780) Armenia. 
\label{fig:xc-types}}
\end{figure*}

For the 14 asteroids with two spectra for each, 12 asteroids are classified as the same type for their two spectra, while two asteroids (696) and (780) received different labels (see Fig. \ref{fig:xc-types}). By the maximum probability that a spectrum belong to a certain type  ( the column 3 of Table \ref{tab:result2}),the extent of variation of spectra of an asteroid, or different asteroids but belong to the same type, could be investigated. As for the two asteroids (696) and (780), we checked the extinction effect on their observations. We found that the airmasses for the two spectra of (696) in the two nights are 1.64 and 1.22, respectively, and that the airmasses of (780) in the same nights are large (2.61 and 2.63). So, we could not rule out the possibility that the diverse labels of the two asteroids rise from the atmospheric effects, although the extinction correction has been done with the averaged extinction curve of the Xinglong station.

\begin{figure*}[ht!]
\centering
\includegraphics[width=0.8\columnwidth]{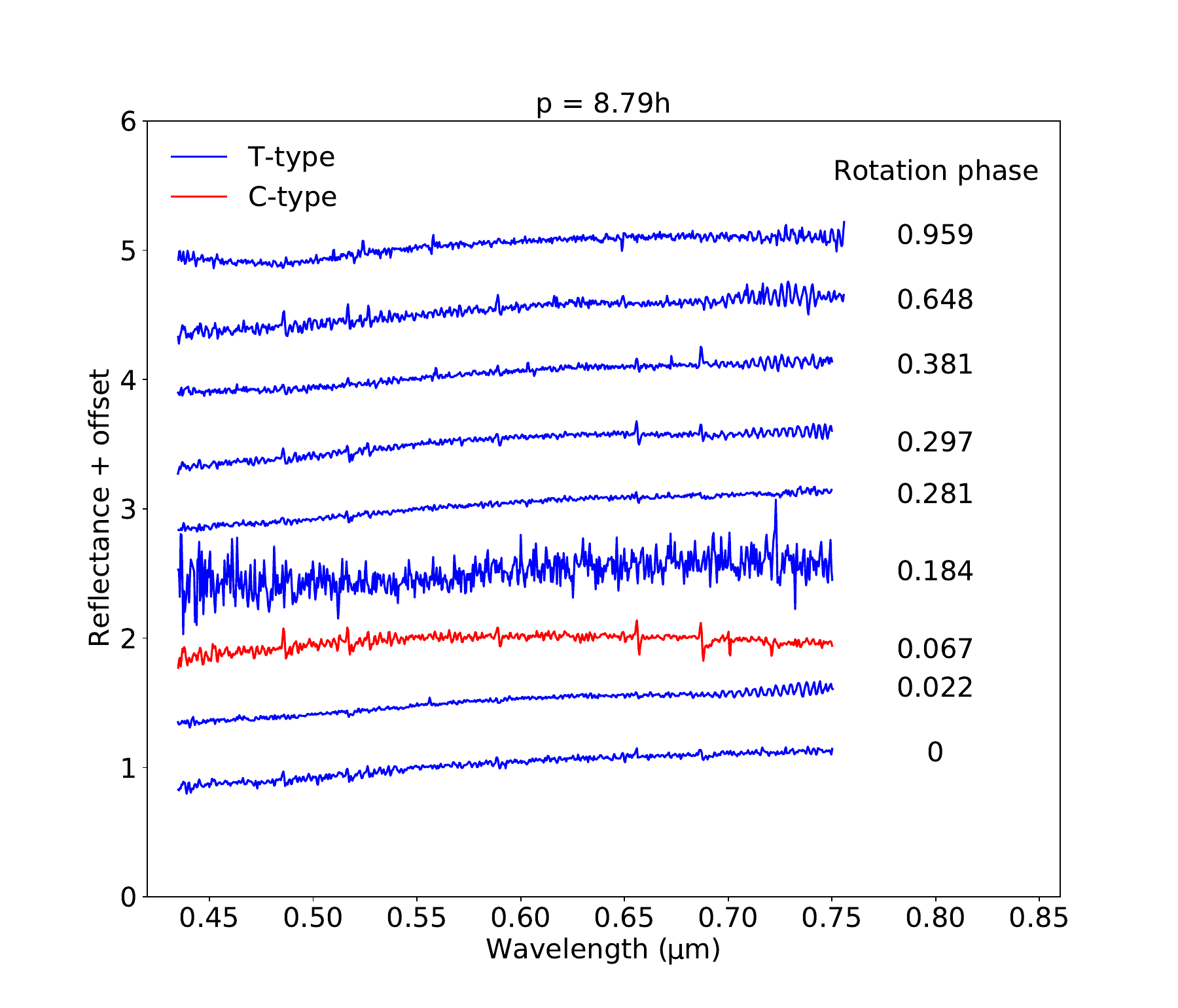}
\caption{Spectra of (469) Argentina. Rightmost column lists the rotational phase folded with a period of 8.79~h. Each spectrum is offset for clarity. 
\label{fig:469-types}}
\end{figure*}

For the asteroid (469) Argentina, we compared its nine spectra obtained in three adjacent nights. Considering the maximum probability of each spectra, eight spectra are labeled as T-type with high probabilities, and one spectrum as C-type (see Table \ref{tab:result2}). The photometric data of Argentina show a more complicated lightcurve shape than that of a sinusoid with two peaks, implying multiple potential periods (17.57~h, 12.84~h, and 8.79~h) of brightness variations \citep{2021MPBu...48...50C,2005EM&P...97..233W,2007MPBu...34...32W}. Here, we selected a spin period of 8.79~h (this period gives a double-peaked lightcurve over a full rotation) to calculate the rotational phase (zero phase at JD2454096.14340). Figure \ref{fig:469-types} shows nine spectra of Argentina arranged from bottom to top with increasing rotational phase angle. From Fig.~\ref{fig:469-types}, the most significant variation occurs at the phase of 0.067~deg, and spectral variations at other rotational phases could be related to the values of maximum probability of each spectrum in the ANN's output layer (see Table \ref{tab:result2}). 

\begin{table}
\caption{Classification results of asteroids with multiple spectra}
\label{tab:result2}
\centering  
\resizebox{\textwidth}{!}{
\begin{tabular}{l l l| l l l}    
\hline\hline   
\renewcommand{\thefootnote}{\fnsymbol{footnote}}
Targets & Observation date & Type and Probability & Targets & Observation date & Type and Probability\\ 
\hline
34 & 2007-11-17.681 & B:74\%  C:24\% &567 & 2007-11-17.577&X:99\% C:0.4\% \\

 & 2007-11-17.702 &B:97\%  C:2\% &   & 2007-11-17.619 &X:90\% C:8\%\\

35 & 2007-11-17.470 & C:71\% K:26\% &613 & 2006-12-26.597&D:99\% T:0.9\%\\

 & 2007-11-17.497 & C:76\% K:19\%  && 2006-12-26.619&D:75\% T:24\%\\

95 & 2007-11-17.783 & C:56\% X:43\%& 680 & 2006-12-28.515 &X:67\% T:22\%\\

 & 2007-11-19.730 & C:88\% X:11\% &  & 2006-12-28.551 &X:68\% K:20\%\\

 199 & 2006-12-28.649 & T:94\% K:3\%& 696 & 2006-12-26.916 &C:99\%\\

 & 2006-12-28.664 & T:86\% K:10\% & &2006-12-27.842 &X:87\% T:9\%\\

 326 & 2006-12-26.546 & X:97\% T:2\%&  772 & 2006-12-27.555&T:89\% X:10\%\\

& 2006-12-27.485 &  X:98\% T:1\% & & 2006-12-27.584 &T:98\% X:1\%\\

381 & 2007-11-17.837 & C:71\% B:28\%  & 780 & 2006-12-27.767 &X:89\% K:7\%\\

& 2007-11-17.872 & C:58\% B:41\%&&2006-12-27.799 &C:91\% X:7\%\\

469 & 2006-12-26.643 &T:94\% K:3\%& 1004 & 2006-12-26.773 &D:99\% T:0.6\%\\

& 2006-12-26.746 & C:99\% B:0.1\%& & 2006-12-28.699 &D:60\% T:39\%\\

& 2006-12-26.668 & T:88\% L:7\%& 1303 & 2006-12-28.437 & D:58\% L:41\%\\

& 2006-12-27.444 & T:96\% K:2\% && 2006-12-28.460 & D:89\% L:10\%\\

& 2006-12-27.516 &  T:82\% K:16\% & \\

& 2006-12-27.615 &  T:92\% K:6\%&\\

& 2006-12-27.729 &  T:98\% K:1\%&\\

& 2006-12-28.486 & T:83\% K:15\% &\\

& 2006-12-28.587 & T:87\% K:6\%&\\

\hline                  
\end{tabular}
}
\end{table}

\section{Summary} \label{sec:sum}

The main aims of our work are to build an ANN tool to classify the future slitless spectra of asteroids from the Chinese Space Survey Telescope (CSST). By the way, we made a taxonomic analysis for 64 unpublished spectra of 42 asteroids with the established ANN tool. The ANN Tool is composed of  five ANNs to give the final result for a tested spectrum by voting.  Each ANN consists of an input layer of 151 neurons, a hidden layer, and an output layer of 10 neurons (i.e., A, B, C, D, T, K, V, L, S, and X-type). Considering the spectral wavelengths of the CSST's spectroscopic instrument,  spectra of 834 asteroids selected from  the SMASS II dataset and their labels from the  Bus-Binzel taxonomic system are applied to train our ANN with different initial weights and biases. 

Following the resolution of the CSST spectroscopic data, spectra of SMASS II are re-sampled by a cubic spline method. To overcome the imbalance of sample numbers in SMASS II between the different categories, we cloned samples for each type until the number of samples of each type reached 400.  Finally, the training data set contains a total of 4,000 spectra. The training procedure applies the mini-batch gradient descent (MBGD) algorithm. The accuracy of each of the five ANNs is higher than 92 \%.

As an application of our ANN tool, 64 unpublished spectra of 42 asteroids are analyzed. 40 out of the 42 asteroids are primitive small objects because they are classified as one of the B, C, X, T, and D-types, and the remaining two are K and S-type. For the first time, asteroid (1303) received a classification: based on our analysis, it is a D-type asteroid.

Comparing our labels of the asteroids to the previous labels, inconsistent labels mainly occur from (1) C-type to X and B-type and from (2) P-type to D and T-type. These cases may be explained with the space weathering effect. \cite{2018Icar..302...10L} found that the space weathering effect could make an original C-type spectrum to be a redder X-type or a bluer B-type depending on the composition of its surface, and a D-type asteroid could be X-type or P-type.

The ANN technique gives us the opportunity to check the spectral variation of an asteroid along the rotational phase. For example, the classification results of 9 spectra of asteroid (469) show uniform composition along its rotational phases except for a single rotational phase indicating C-type. 

By the ANN analysis, 40 primitive asteroids are classifed as B, C, X, T, or D-type, reflecting the diversity of primitive asteroids. Investigating visual slopes of 40 asteroids, we found that slopes (a linear fit to the spectrum between $0.55$--$0.86$~\textmu{}m) of  B, X, T, and D-types are located in $-3.5$--$0.11$, $1.5$--$4.5$, $4.6$--$6.3$, and $4.9$--$9.9$ \%/1000\AA, respectively,  which is consistent with the trend of the  visual slope (also the dominated feature) of these four types. As for the C-type asteroids, wide visual slopes are presented. In detail, most values occur between $-0.82$--$1.64$, while 5 spectra (see Fig.~\ref{fig:all} for 146, 381-2, 414, and 508; and Fig.~\ref{fig:469-types} for (469)) have more negative slopes. When checking these five spectra, a turning point around 0.5~\textmu{}m and an absorption band around 0.7~\textmu{}m can be discerned, which imply hydrous minerals on the asteroids' surfaces \citep{2018Icar..302...10L,2018Icar..311...35D}.

To identify the feature of 0.7~\textmu{}m absorption band in a spectrum, linear trend of the spectrum derived by fitting  two section data $0.55$--$0.58$~\textmu{}m and $0.83$--$0.86$~\textmu{}m is removed. Here, we think, a possible absorption band feature must be a concave shape with a maximum depth deeper than 1 percent of values of two shoulder parts. With this idea, other observed spectra are checked, and the results are 10 out of 12 spectra of B-type, 9 out of 17 spectra of C-type (including 5 spectra mentioned above), and 2 D-type and 1 T-type show this absorption band feature. It seems that the proportion of B-type asteroids bearing hydrous minerals is the largest, followed by the C-type.

We are aware that the presented ANN tool used to analyse our spectral data needs to be improved for satisfying the future needs for the spectral data of asteroids from the CSST survey. This is because the new data will contain important UV data ($0.255$--$0.43$~\textmu{}m) and data from $0.83$--$1.0$~\textmu{}m, which are keys to understand  primordial objects in the Solar System. More importantly, we will obtain spectroscopic data of numerous faint Solar System objects from the CSST ten years survey, which may improve significantly our understanding to the origin and evolution of small bodies of the Solar System, and even planetary systems.

\section{Acknowledgements}

This work is supported by the National Natural Science Foundation of China (grants No. 11673063 and 12373069 ) and the Research Council of Finland (grants No. 345115 and 336546). We acknowledge the science research grants from the China Manned Space Project with No. CMS-CSST-2021-B08, the Foreign Experts Project (FEP) of State Administration of Foreign Experts Affairs of China (SAFEA) with No. G2021039001L, and the Chinese Academy of Sciences President’s International Fellowship Initiative (PIFI) Grant No. 2021VMA0017.

\bibliography{sample631}{}

\begin{thebibliography}{}
\expandafter\ifx\csname natexlab\endcsname\relax\def\natexlab#1{#1}\fi
\providecommand{\url}[1]{\href{#1}{#1}}
\providecommand{\dodoi}[1]{doi:~\href{http://doi.org/#1}{\nolinkurl{#1}}}
\providecommand{\doeprint}[1]{\href{http://ascl.net/#1}{\nolinkurl{http://ascl.net/#1}}}
\providecommand{\doarXiv}[1]{\href{https://arxiv.org/abs/#1}{\nolinkurl{https://arxiv.org/abs/#1}}}

\bibitem[{{Ball} \& {Brunner}(2010)}]{2010IJMPD..19.1049B}
{Ball}, N.~M., \& {Brunner}, R.~J. 2010, International Journal of Modern
  Physics D, 19, 1049, \dodoi{10.1142/S0218271810017160}

\bibitem[{{Baron}(2019)}]{2019arXiv190407248B}
{Baron}, D. 2019, arXiv e-prints, arXiv:1904.07248.
\newblock \doarXiv{1904.07248}

\bibitem[{{Belskaya} {et~al.}(2017){Belskaya}, {Fornasier}, {Tozzi},
  {Gil-Hutton}, {Cellino}, {Antonyuk}, {Krugly}, {Dovgopol}, \&
  {Faggi}}]{2017Icar..284...30B}
{Belskaya}, I.~N., {Fornasier}, S., {Tozzi}, G.~P., {et~al.} 2017, \icarus,
  284, 30, \dodoi{10.1016/j.icarus.2016.11.003}

\bibitem[{Boer {et~al.}(2005)Boer, Kroese, Mannor, \& Rubinstein}]{2005A}
Boer, P. T.~D., Kroese, D.~P., Mannor, S., \& Rubinstein, R.~Y. 2005, Annals of
  Operations Research, 134, 19

\bibitem[{{Bottke} {et~al.}(2002){Bottke}, {Cellino}, {Paolicchi}, \&
  {Binzel}}]{2002aste.book.....B}
{Bottke}, W.~F., J., {Cellino}, A., {Paolicchi}, P., \& {Binzel}, R.~P. 2002,
  {Asteroids III} (University of Arizona Press)

\bibitem[{{Bus} \& {Binzel}(2002{\natexlab{a}})}]{2002Icar..158..146B}
{Bus}, S.~J., \& {Binzel}, R.~P. 2002{\natexlab{a}}, \icarus, 158, 146,
  \dodoi{10.1006/icar.2002.6856}

\bibitem[{{Bus} \& {Binzel}(2002{\natexlab{b}})}]{2002Icar..158..106B}
---. 2002{\natexlab{b}}, \icarus, 158, 106, \dodoi{10.1006/icar.2002.6857}

\bibitem[{{Bus} {et~al.}(2002){Bus}, {Vilas}, \&
  {Barucci}}]{2002aste.book..169B}
{Bus}, S.~J., {Vilas}, F., \& {Barucci}, M.~A. 2002, in Asteroids III
  (University of Arizona Press), 169--182

\bibitem[{Clark {et~al.}(2002)Clark, Helfenstein, Iii, Peterson, Veverka,
  Izenberg, Domingue, Wellnitz, \& Mcfadden}]{2002NEAR}
Clark, B.~E., Helfenstein, P., Iii, J., {et~al.} 2002, Icarus, 155, 189

\bibitem[{{Colazo} {et~al.}(2022){Colazo}, {Alvarez-Candal}, \&
  {Duffard}}]{2022arXiv220405075C}
{Colazo}, M., {Alvarez-Candal}, A., \& {Duffard}, R. 2022, arXiv e-prints,
  arXiv:2204.05075.
\newblock \doarXiv{2204.05075}

\bibitem[{{Colazo} {et~al.}(2021){Colazo}, {Stechina}, {Fornari}, {Santucho},
  {Mottino}, {Pulver}, {Melia}, {Su{\'a}rez}, {Scotta}, {Chapman}, {Oey},
  {Meza}, {Bellocchio}, {Morales}, {Speranza}, {Romero}, {Suligoy},
  {Passarino}, {Borello}, {Farf{\'a}n}, {Lim{\'o}n}, {Delgado}, {Naves}, \&
  {Colazo}}]{2021MPBu...48...50C}
{Colazo}, M., {Stechina}, A., {Fornari}, C., {et~al.} 2021, Minor Planet
  Bulletin, 48, 50

\bibitem[{{De Pr{\'a}} {et~al.}(2018){De Pr{\'a}}, {Pinilla-Alonso}, {Carvano},
  {Licandro}, {Campins}, {Moth{\'e}-Diniz}, {De Le{\'o}n}, \&
  {Al{\'\i}-Lagoa}}]{2018Icar..311...35D}
{De Pr{\'a}}, M.~N., {Pinilla-Alonso}, N., {Carvano}, J.~M., {et~al.} 2018,
  \icarus, 311, 35, \dodoi{10.1016/j.icarus.2017.11.012}

\bibitem[{{DeMeo} {et~al.}(2009){DeMeo}, {Binzel}, {Slivan}, \&
  {Bus}}]{2009Icar..202..160D}
{DeMeo}, F.~E., {Binzel}, R.~P., {Slivan}, S.~M., \& {Bus}, S.~J. 2009,
  \icarus, 202, 160, \dodoi{10.1016/j.icarus.2009.02.005}

\bibitem[{{DeMeo} \& {Carry}(2014)}]{2014Natur.505..629D}
{DeMeo}, F.~E., \& {Carry}, B. 2014, \nat, 505, 629,
  \dodoi{10.1038/nature12908}

\bibitem[{Goodfellow {et~al.}(2016)Goodfellow, Bengio, \&
  Courville}]{Goodfellow-et-al-2016}
Goodfellow, I., Bengio, Y., \& Courville, A. 2016, Deep Learning (MIT Press)

\bibitem[{{Howell} {et~al.}(1994){Howell}, {Merenyi}, \&
  {Lebofsky}}]{1994JGR....9910847H}
{Howell}, E.~S., {Merenyi}, E., \& {Lebofsky}, L.~A. 1994, \jgr, 99, 10847,
  \dodoi{10.1029/93JE03575}

\bibitem[{{Huang} {et~al.}(2017){Huang}, {Ma}, {Zhao}, \&
  {Lu}}]{2017ChA&A..41..549H}
{Huang}, C., {Ma}, Y.-h., {Zhao}, H.-b., \& {Lu}, X.-p. 2017, \caa, 41, 549,
  \dodoi{10.1016/j.chinastron.2017.11.006}

\bibitem[{{Ivezi{\'c}} {et~al.}(2019){Ivezi{\'c}}, {Connelly}, {Vanderplas}, \&
  {Gray}}]{2019sdmm.book.....I}
{Ivezi{\'c}}, {\v{Z}}., {Connelly}, A.~J., {Vanderplas}, J.~T., \& {Gray}, A.
  2019, {Statistics, Data Mining, and Machine Learning in Astronomy} (Princeton
  University Press)

\bibitem[{{Jewitt}(2002)}]{2002AJ....123.1039J}
{Jewitt}, D.~C. 2002, \aj, 123, 1039, \dodoi{10.1086/338692}

\bibitem[{{Klimczak} {et~al.}(2021){Klimczak}, {Kot{\l}owski}, {Oszkiewicz},
  {DeMeo}, {Kryszczy{\'n}ska}, {Wilawer}, \& {Carry}}]{2021FrASS...8..216K}
{Klimczak}, H., {Kot{\l}owski}, W., {Oszkiewicz}, D., {et~al.} 2021, Frontiers
  in Astronomy and Space Sciences, 8, 216, \dodoi{10.3389/fspas.2021.767885}

\bibitem[{{Lantz} {et~al.}(2018){Lantz}, {Binzel}, \&
  {DeMeo}}]{2018Icar..302...10L}
{Lantz}, C., {Binzel}, R.~P., \& {DeMeo}, F.~E. 2018, \icarus, 302, 10,
  \dodoi{10.1016/j.icarus.2017.11.010}

\bibitem[{{Lazzaro} {et~al.}(2007){Lazzaro}, {Angeli}, {Carvano},
  {Mothe-Diniz}, {Duffard}, \& {Florczak}}]{2007PDSS...51.....L}
{Lazzaro}, D., {Angeli}, C.~A., {Carvano}, J.~M., {et~al.} 2007, NASA Planetary
  Data System, EAR

\bibitem[{{Mahlke} {et~al.}(2022){Mahlke}, {Carry}, \&
  {Mattei}}]{2022arXiv220311229M}
{Mahlke}, M., {Carry}, B., \& {Mattei}, P.-A. 2022, arXiv e-prints,
  arXiv:2203.11229.
\newblock \doarXiv{2203.11229}

\bibitem[{{Penttil{\"a}} {et~al.}(2021){Penttil{\"a}}, {Hietala}, \&
  {Muinonen}}]{2021A&A...649A..46P}
{Penttil{\"a}}, A., {Hietala}, H., \& {Muinonen}, K. 2021, \aap, 649, A46,
  \dodoi{10.1051/0004-6361/202038545}

\bibitem[{{Popescu} {et~al.}(2018){Popescu}, {Licandro}, {Carvano},
  {Stoicescu}, {de Le{\'o}n}, {Morate}, {Boac{\u{a}}}, \&
  {Cristescu}}]{2018A&A...617A..12P}
{Popescu}, M., {Licandro}, J., {Carvano}, J.~M., {et~al.} 2018, \aap, 617, A12,
  \dodoi{10.1051/0004-6361/201833023}

\bibitem[{{Tholen}(1984)}]{1984atca.book.....T}
{Tholen}, D.~J. 1984, {Asteroid taxonomy from cluster analysis of photometry}
  (The University of Arizona)

\bibitem[{{Wang} {et~al.}(2005){Wang}, {Zhang}, \& {Gu}}]{2005EM&P...97..233W}
{Wang}, X.-B., {Zhang}, X.-L., \& {Gu}, S.-H. 2005, Earth Moon and Planets, 97,
  233, \dodoi{10.1007/s11038-006-9094-6}

\bibitem[{{Warner}(2007)}]{2007MPBu...34...32W}
{Warner}, B.~D. 2007, Minor Planet Bulletin, 34, 32

\bibitem[{{Zellner} {et~al.}(1985){Zellner}, {Tholen}, \&
  {Tedesco}}]{1985Icar...61..355Z}
{Zellner}, B., {Tholen}, D.~J., \& {Tedesco}, E.~F. 1985, \icarus, 61, 355,
  \dodoi{10.1016/0019-1035(85)90133-2}

\bibitem[{Zhan(2021)}]{zhanhu}
Zhan, H. 2021, Chinese Science Bulletin, 66, 1290

\end{thebibliography}
\bibliographystyle{aasjournal}

\end{document}